\documentclass{aa}

\usepackage{graphicx}
\usepackage{txfonts}
\usepackage{arydshln}
\usepackage[colorlinks=true,citecolor=blue,urlcolor=blue]{hyperref}
\usepackage{url}
\usepackage[usenames,dvipsnames]{xcolor}
\usepackage{booktabs}
\usepackage[utf8]{inputenc}

\newcommand{\phz}{photo-$z$}
\newcommand{\phzs}{photo-$z$s}
\newcommand{\spz}{spectro-$z$}
\newcommand{\spzs}{spectro-$z$s}
\newcommand{\zsp}{z_\mathrm{spec}}
\newcommand{\zph}{z_\mathrm{phot}}
\newcommand{\meanz}{\langle z \rangle}
\newcommand{\meandz}{\langle \delta z \rangle}
\newcommand{\ttt}{\texttt}
\newcommand{\mtt}{\mathtt}
\newcommand{\mrm}{\mathrm}
\defcitealias{Bilicki18}{B18}

\begin{document} 

    \title{Bright galaxy sample in the Kilo-Degree Survey Data Release 4}
\subtitle{Selection, photometric redshifts, and physical properties}
\titlerunning{KiDS-Bright galaxy sample}

\author{
M.~Bilicki\inst{\ref{Warsaw} \thanks{\email{\url{bilicki@cft.edu.pl}}}  } \and
A.~Dvornik\inst{\ref{Bochum},\ref{Leiden}
\thanks{\email{\url{dvornik@astro.ruhr-uni-bochum.de}}} } \and
H.~Hoekstra\inst{\ref{Leiden}
\thanks{\email{\url{hoekstra@strw.leidenuniv.nl}}} } \and
A.H.~Wright\inst{\ref{Bochum}}\and
N.E.~Chisari\inst{\ref{Utrecht}} \and
M.~Vakili\inst{\ref{Leiden}} \and\\
M.~Asgari\inst{\ref{Edinburgh}} \and
B.~Giblin\inst{\ref{Edinburgh}} \and
C.~Heymans\inst{\ref{Edinburgh},\ref{Bochum}} \and
H.~Hildebrandt\inst{\ref{Bochum}}\and
B.W.~Holwerda\inst{\ref{Louisville}}\and
A.~Hopkins\inst{\ref{Macquarie}}\and
H.~Johnston\inst{\ref{Utrecht}}\and
A.~Kannawadi\inst{\ref{Princeton}}\and
K.~Kuijken\inst{\ref{Leiden}}\and
S.J.~Nakoneczny\inst{\ref{NCBJ}}\and
H.Y.~Shan\inst{\ref{Shanghai},\ref{Beijing}} \and
A.~Sonnenfeld\inst{\ref{Leiden}} \and
E.~Valentijn\inst{\ref{Groningen}}
}

\institute{
Center for Theoretical Physics, Polish Academy of Sciences, al. Lotnik\'{o}w 32/46, 02-668 Warsaw, Poland \label{Warsaw} 
\and
Ruhr University Bochum, Faculty of Physics and Astronomy, Astronomical Institute (AIRUB), German Centre for Cosmological Lensing, 44780 Bochum, Germany \label{Bochum} 
\and
Leiden Observatory, Leiden University, P.O. Box 9513, NL-2300 RA Leiden, The Netherlands \label{Leiden}
\and
Institute for Theoretical Physics, Utrecht University, Princetonplein 5, 3584 CC Utrecht, The Netherlands \label{Utrecht}
\and
Institute for Astronomy, University of Edinburgh, Blackford Hill, Edinburgh, EH9 3HJ, UK \label{Edinburgh}
\and
Department of Physics and Astronomy, University of Louisville, 102 Natural Science Building, Louisville KY 40292, USA \label{Louisville}
\and
 Australian Astronomical Optics, Macquarie University, 105 Delhi Rd, North Ryde, NSW 2113, Australia  \label{Macquarie}
\and
Department of Astrophysical Sciences, Princeton University, 4 Ivy Lane, Princeton, NJ 08544, USA \label{Princeton}
\and
National Centre for Nuclear Research, Astrophysics Division, ul. Pasteura 7, 02-093 Warsaw, Poland \label{NCBJ}
\and
Shanghai Astronomical Observatory (SHAO), Nandan Road 80, Shanghai 200030, China \label{Shanghai}
\and
University of Chinese Academy of Sciences, Beijing 100049, China \label{Beijing}
\and
Kapteyn Institute, University of Groningen, PO Box 800, NL 9700 AV Groningen, The Netherlands \label{Groningen}
}

\authorrunning{Bilicki, Dvornik, Hoekstra et al.}

\date{Received 15 January 2021 / Accepted 24 May 2021}

\abstract{
We present a bright galaxy sample with accurate and precise photometric redshifts (\phzs), selected using $ugriZYJHK_\mrm{s}$ photometry from the Kilo-Degree Survey (KiDS) Data Release 4 (DR4). The highly pure and complete dataset is flux-limited at $r<20$ mag, covers $\sim1000$ deg$^2$, and contains about 1 million galaxies after artifact masking. We exploit the overlap with Galaxy And Mass Assembly (GAMA) spectroscopy as calibration to determine \phzs\ with the supervised machine learning neural network algorithm implemented in the ANNz2 software. The \phzs\ have a mean error of $|\meandz| \sim 5 \times 10^{-4}$ and low scatter (scaled mean absolute deviation of $\sim 0.018(1+z)$); they are both practically independent of the $r$-band magnitude and \phz\ at $0.05 < \zph < 0.5$. Combined with the 9-band photometry, these allow us to estimate robust absolute magnitudes and stellar masses for the full sample.
As a demonstration of the usefulness of these data, we split the dataset into red and blue galaxies, used them as lenses, and measured the weak gravitational lensing signal around them for five stellar mass bins. We fit a halo model to these high-precision measurements to constrain the stellar-mass--halo-mass relations for blue and red galaxies. We find that for high stellar mass ($M_\star>5\times 10^{11} M_\odot$), the red galaxies occupy dark matter halos that are much more massive than those occupied by blue galaxies with the same stellar mass. The data presented here are publicly released via the KiDS webpage at \url{http://kids.strw.leidenuniv.nl/DR4/brightsample.php}.
}


\maketitle 


\section{Introduction}

Galaxies are not distributed randomly throughout the Universe: they trace the underlying dark matter distribution, which itself forms a web-like structure under the influence of gravity in an expanding universe. For a given cosmological model, the growth of a structure can be simulated using cosmological numerical simulations, and
the statistical properties of the resulting matter distribution as a function of scale and redshift can thus be robustly predicted. Given a prescription that relates their properties to the matter distribution, the observed spatial distribution of galaxies can thus be used to infer cosmological parameter estimates \citep[e.g.,][]{Percival01,Cole05,Alam17,eBOSS}.

The galaxy redshift is a key observable in such analyses, and large spectroscopic surveys have therefore played an important role in establishing the current cosmological model. For large-scale clustering studies, it is advantageous to target specific subsets of galaxies rather sparsely because the survey can cover larger areas more efficiently. Consequently, most current results are based on redshift surveys that target specific galaxy types, such as luminous red galaxies \citep[LRGs;][]{BOSS,2dFLenS}. The downside of such strategies, however, is that detailed information about the environment is typically lost. 

In contrast, a highly complete spectroscopic survey can only cover relatively small areas because fiber collisions or slit overlaps prevent or limit simultaneous spectroscopy of close galaxies; repeat visits are required to achieve a high completeness. For studies of galaxy formation and evolution, this can nonetheless be fruitful, as the Galaxy And Mass Assembly survey \citep[GAMA,][]{GAMA} has demonstrated
\citep[e.g.,][]{Gunawardhana11,Robotham11,Baldry12}. Although many of these applications rely on spectroscopic redshifts, several questions can still be addressed with less precise (photometric) redshift information over large areas.

To study the connection between galaxy properties and the dark matter distribution around galaxies, weak gravitational lensing has become an important observational tool. The foreground galaxies, which are embedded in dark matter dominated halos, act as lenses that distort space-time around them, leading to correlations in the shapes of more distant galaxies. This so-called (weak) galaxy-galaxy lensing (GGL) is used to study the stellar-mass--halo-mass relation \citep[e.g.,][]{Leauthaud12, Coupon15, vanUitert16} to examine the galaxy bias \citep[e.g.,][]{Hoekstra02, Dvornik18}, or to test modified gravity theories \citep[e.g.,][]{Tian09, Brouwer17}. Combined with measurements of the clustering of galaxies and the cosmic shear signal, so-called 3$\times$2pt analyses provide competitive constraints on cosmological parameters \citep[e.g.,][]{Abbott18,Joudaki18,vanUitert18,Heymans20}.
These applications rely on an overlapping sample of lenses with precise redshifts and a background sample with a large number of distant sources with reliable shape measurements. The latter are improving thanks to large, deep, multiband imaging surveys that cover increasingly larger areas of the sky, with the aim of measuring cosmological parameters using weak gravitational lensing, such as the Kilo-Degree Survey \citep[KiDS,][]{KiDS}, the Dark Energy Survey \citep[DES,][]{DES}, and the Hyper-Suprime Cam Subaru Strategic Program \citep{HSC}.

In this paper we focus on KiDS, which covers 1350 deg$^2$ in nine broadband filters at optical and near-infrared (NIR) wavelengths. Unfortunately, the spectroscopic samples that overlap with the survey only yield $\sim110$ lenses per square degree in the case of the Baryon Oscillation Spectroscopic Survey \citep[BOSS,][]{BOSS}, and $\sim40$ deg$^{-2}$ for the 2-degree Field Lensing Survey \citep[2dFLenS,][]{2dFLenS}. They jointly cover the full final KiDS area of 1350 deg$^2$, and they have been exploited to test general relativity \citep{Amon18,Blake20} and to constrain cosmological parameters \citep{Joudaki18,Heymans20,Troster20}, but their low number density limits the range of applications.

In contrast, GAMA provides a much denser sampling of up to 1000 lenses per deg$^2$ (albeit at a lower mean redshift than BOSS or 2dFLenS), allowing for unique studies of the lensing signal as a function of environment \citep[e.g.,][]{Sifon15,Viola15,Brouwer16,vanUitert17,Linke20}, but its overlap with KiDS is limited to $\sim230$ deg$^2$.
Hence for studies of the small-scale lensing signal, or studies of galaxies other than LRGs, we cannot rely on spectroscopic-only coverage over the full KiDS survey area. Fortunately, for many applications less precise photometric redshifts (\phzs) suffice \citep[e.g.,][]{Brouwer18}, provided that the actual lens redshift distribution is accurately known.

In \citet[][\citetalias{Bilicki18} hereafter]{Bilicki18}, we used the third KiDS data release \citep[DR3,][]{KiDS-DR3} covering 450 deg$^2$ and showed that by applying a limit of $r\lesssim 20$ to the imaging data, it was possible to extract a galaxy sample with a surface number density of $\sim1000$ deg$^{-2}$ at a mean redshift $\meanz=0.23$. Taking advantage of the overlap with GAMA spectroscopy, and using optical-only photometry ($ugri$) available from KiDS DR3, we obtained \phzs\ that had a negligible bias with $\meandz \sim  10^{-4}$ and a small scatter of $\sigma_{\delta z / (1+z)}\sim0.022$. 
These redshift statistics were achieved by deriving \phzs\ using a supervised machine-learning (ML) artificial neural networks (ANN) algorithm \citep[ANNz2,][]{ANNz2}, which was trained on galaxies
with spectroscopic redshifts (\spzs) in common between KiDS and GAMA. Such a good \phz\ performance was possible thanks to the very high spectroscopic completeness of GAMA in its three equatorial fields (G09, G12, \& G15): at the limit of $r<19.8$, only $\sim1.5\%$ of the targets (preselected from SDSS) do not have a spectroscopic redshift measured there \citep{GAMA-II}. As GAMA is essentially a complete subset of the much deeper KiDS dataset, restricting the latter to the flux limit of the former allowed us to take full advantage of the main supervised ML benefit: if a well-matched training set is available, then \phzs\ derived with this technique is accurate and precise. 

Here we extend the successful analysis of \citetalias{Bilicki18} to a larger area and broader wavelength coverage using the imaging data from the fourth public KiDS data release \citep[DR4;][]{KiDS-DR4}. We improve upon the earlier results and derive statistically precise and accurate \phzs\ for a flux-limited sample of bright galaxies without any color preselection. The imaging data cover about 1000\,deg$^2$ in nine filters, combining KiDS optical photometry with NIR data from the VISTA Kilo-degree Infrared Galaxy survey \citep[VIKING,][]{VIKING}. As shown in
\citetalias{Bilicki18}, the addition of the NIR data should improve the \phz\ performance with respect to the earlier work.
Following that previous study, we take advantage of the  overlapping spectroscopy from GAMA, which allows for a robust empirical calibration. This leads to better individual redshift estimates for bright, low redshift galaxies, both in terms of lower bias and reduced scatter, compared to the default \phz\ estimates that are provided as part of KiDS DR4. Those \phzs\ were derived with the Bayesian Photometric Redshift approach \citep[BPZ;][]{BPZ}, with settings optimized for relatively faint ($r>20$) and high-$z$ cosmic shear sources, which makes them suboptimal for bright, low-redshift galaxies \citep[\citetalias{Bilicki18};][]{Vakili19}.

Over the full KiDS DR4 footprint of $\sim1000$ deg$^2$, we selected a flux-limited galaxy sample, closely matching the GAMA depth ($r<20$), and we derived \phzs\ for all the  objects with 9-band detections. We call this sample KiDS-Bright for short. The final catalog includes about a million galaxies after artifact masking, that is $\sim 1000$ objects per square degree. The inclusion of the NIR photometry reduces the \phz\ scatter to $\sigma_{\delta z / (1+z)}\sim0.018$, whilst still retaining a very small bias of $|\meandz| <  10^{-3}$.

As a further extension of the previous results (\citetalias{Bilicki18}), we derived absolute magnitudes and stellar masses for the KiDS-Bright sample, using the {\sc LePhare} \citep{Arnouts99,Ilbert06} spectral energy distribution fitting software. As an example of a scientific application of this dataset, we present a study of the stellar-to-halo-mass relation using GGL,
where we split the sample into blue and red galaxies.

This paper is organized as follows. In Sect.~\ref{Sec:data} we describe the data used: KiDS in Sect.~\ref{Sec:KiDS_data}, GAMA in Sect.~\ref{Sec:GAMA}, and the selection of the KiDS-Bright sample in Sect.~\ref{Sec:KiDS-bright-selection}. In Sect.~\ref{Sec:photo-zs} we present the photometric redshift estimation, quantify the \phz\ performance (Sect.~\ref{sec:photo-z-performance}), and provide a model for redshift errors (Sect.~\ref{Sec:photo-z-model}). In Sect.~\ref{Sec:stellar_masses} we discuss the stellar mass and absolute magnitude derivation, validate it with GAMA, and provide details of the red and blue galaxy selection. We present the GGL measurements using this sample in Sect.~\ref{Sec:GGL}, compare them to the signal from GAMA in Sect.~\ref{sec:GGLGAMA}, and use them to constrain the stellar-to-halo mass relation in Sect.~\ref{Sec:SHMR}. We conclude in Sect.~\ref{Sec:conclusions_and_prospects}.

The paper is accompanied by the public release of the data presented here\footnote{Data will be available upon publication. Please contact the authors for earlier access.}, including the \phzs\ and estimates of physical properties for the full KiDS-Bright galaxy sample over the $\sim1000$ deg$^2$ footprint of KiDS DR4. 

\section{Data and sample selection}
\label{Sec:data}

\subsection{KiDS imaging data}
\label{Sec:KiDS_data}

To select our galaxy sample, we used photometry in nine bands
from a joint analysis of KiDS ($ugri$) and VIKING ($ZYJHK_\mrm{s}$) data that form the fourth public KiDS data release \citep{KiDS-DR4}\footnote{See \url{http://kids.strw.leidenuniv.nl/DR4/index.php} for data access.}. This combined data set, which we refer to as "KV", covers an area of approximately $1000$ deg$^2$, limited by the KiDS 4-band observations obtained by January 24, 2018 (VIKING had fully finished earlier). KiDS imaging was obtained with the OmegaCAM camera \citep{Kuijken11} at the VLT Survey Telescope \citep{Capaccioli12}, while VIKING employed the VIRCAM \citep{VIRCAM} on the Visible and Infrared Survey Telescope for Astronomy \citep[VISTA,][]{VISTA}.

The imaging data were processed using dedicated pipelines: the Astro-WISE information system \citep{AW} for the production of co-added images ("coadds") in the four optical bands, and a \textsc{theli} \citep{THELI} $r$-band image reduction to provide a source catalog suitable for the core weak lensing science case. The VIKING magnitudes for KiDS DR4 were obtained from forced photometry on the \textsc{theli}-detected sources, using a re-reduction of the NIR imaging that started from the VISTA ``paw-prints'' processed by the Cambridge Astronomical Survey Unit (CASU).

Photometric redshift estimates rely on robust colors, for which we use the Gaussian Aperture and Photometry \citep[\textsc{GAaP},][]{GAaP} measurements, which in DR4 are provided for all the bands. They are obtained via a homogenization procedure in which calibrated and stacked images are first "Gaussianized," that is the point-spread-function (PSF) is homogenized across each individual coadd.  The photometry is then measured using a Gaussian-weighted aperture (based on the $r$-band ellipticity and orientation) that compensates for seeing differences between the different filters (see \citealt{KiDS-GL} for more details). 
Our ML \phz\ derivation requires that magnitudes are available in all of the filters employed. Hence we require that the sources have data and detections in all nine bands. 

The \textsc{GAaP} magnitudes are useful for accurate color estimates, but they miss part of the flux for extended sources. Various other magnitude estimates are, however, provided for the $r$-band data. Here we use the 
Kron-like automatic aperture \ttt{MAG\_AUTO} and the isophotal magnitude \ttt{MAG\_ISO}, as measured by {\sc SExtractor} \citep{SExtractor}. 
These are not corrected for Galactic extinction and zero-point variations between different KiDS tiles (unlike the published \textsc{GAaP} magnitudes). To account for this, we define  $r^\mrm{KiDS}_\mrm{auto} = \mtt{MAG\_AUTO} + \mtt{DMAG} - \mtt{EXTINCTION\_R}$  (and analogously for \ttt{MAG\_ISO}), where \ttt{DMAG} are per-tile zero-point offset corrections, and the Galactic extinction at the object position was derived from the \cite{SFD} maps with the \cite{SF11} coefficients. Where unambiguous, we do not use the "KiDS" superscript.

In order to separate galaxies from stars, we used three star-galaxy separation indicators provided in the KiDS DR4 multiband dataset. The first one is the continuous \ttt{CLASS\_STAR} derived with {\sc SExtractor}, ranging from 0 (extended) to 1 (point sources). The second separator is the discrete \ttt{SG2DPHOT} classification bitmap based on the $r$-band detection image source morphology \citep[e.g.,][]{KiDS-DR2}, which for instance is set to 0 for galaxies and 1, 4, or 5 for stars. Lastly, also \ttt{SG\_FLAG} is a discrete star-galaxy separator that is equal to 0 for high-confidence stars and 1 otherwise\footnote{See \cite{KiDS-GL} sect.~3.2.1 for a description of this star-galaxy separation.}. 

The catalogs contain two flags that can be used to identify problematic sources (artifacts). The first one is \ttt{IMAFLAGS\_ISO}, a bitmap of mask flags indicating the types of masked areas that intersect with the isophotes of each source, as identified by the {\sc Pulecenella} software \citep{KiDS-DR2}. We require this flag to be 0. The second flag is the KV multiband bit-wise \ttt{MASK}, which combines Astro-WISE and \textsc{theli} flags for the KiDS and VIKING bands\footnote{See \url{http://kids.strw.leidenuniv.nl/DR4/format.php\#masks} for details.}. It indicates issues with source extraction, such as star halos, globular clusters, saturation, and chip gaps. The recommended selection in DR4 is to remove sources with $(\mtt{MASK} \& 28668)>0$. We did not apply this mask by default in the final dataset, but instead we provide a binary flag indicating whether an object meets this masking criterion or not.

In Section~\ref{Sec:GGL} we explain how we measured the lensing signal around our sample of bright galaxies using shape measurements that are based on the $r$-band images. The galaxy shapes were measured using \emph{lens}fit \citep{lensfit}, which was calibrated with image simulations described in \citet{Kannawadi19}. Those are complemented with \phz\ estimates based on an implementation of the BPZ code \citep{BPZ}. For further details on the 
image reduction, \phz\ calibration, and shape measurement analysis for these background sources, we refer the interested reader to \cite{KiDS-DR4,Giblin20}, and \cite{Hildebrandt20b}. 

\subsection{GAMA spectroscopic data}
\label{Sec:GAMA}

The Galaxy And Mass Assembly survey \citep{GAMA} is a unique spectroscopic redshift and multiwavelength photometric campaign, which employed the AAOmega spectrograph on the Anglo-Australian Telescope to measure galaxy spectra in five fields with a total $\sim286$ deg$^2$ area. Four of these fields (equatorial G09, G12, and G15 of 60 deg$^2$ each, and Southern G23 of $\sim51$ deg$^2$) fully overlap with KiDS, and we exploited
this to optimize the bright galaxy selection and calibrate the \phzs.
Unique features of GAMA include the panchromatic imaging, spanning almost the entire electromagnetic spectrum \citep{Driver16,GAMA-LAMBDAR}, and the detailed redshift sampling in its equatorial fields: it is $98.5\%$ complete for SDSS-selected galaxies with $r<19.8$ mag, providing an almost volume-limited selection at $z \lesssim 0.2$, and it includes a sizable number of galaxies up to $z \sim 0.5$.

In our work we use the "GAMA II" galaxy dataset \citep{GAMA-II} from the equatorial fields, which includes, but is not limited to, the first three public GAMA data releases. The GAMA targets for spectroscopy were selected there from SDSS DR7 imaging \citep{SDSS.DR7}, requiring a \cite{Petro} magnitude $r_\mrm{Petro}<19.8$. Only extended sources were targeted, primarily based on the value of $\Delta_\mrm{sg} = r_\mrm{psf} - r_\mrm{model}$ \citep{SDSS.MGS}, where the two latter magnitudes are the SDSS PSF and model $r$-band measurements, respectively. To improve the point source removal further, the $J-K$ NIR color from the UKIRT Infrared Deep Sky Survey \citep[UKIDSS,][]{UKIDSS} was also used \citep{GAMA-input}. 

In the equatorial fields, GAMA also includes sources fainter than $r=19.8$ and/or selected differently than the main flux limited sample ("filler" targets; see \cite{GAMA-input,GAMA-II,GAMA-DR3} for details). We used these in the KiDS \phz\ training together with the flux-limited sample, but not to calibrate the bright-end selection. KiDS also overlaps with the southern G23 field, but the targets there were selected at a brighter limit ($i<19.2$) than in the equatorial areas, and they were observed at a lower completeness. We therefore did not use that field for our sample selection and \phz\ calibration.  

We used the equatorial fields of GAMA TilingCatv46, which cover roughly 180 deg$^2$ fully within the KiDS DR4 footprint. To ensure robust spectroscopy, we required a redshift quality\footnote{GAMA galaxies with $\mathtt{NQ} < 3$ are often fainter than the completeness limit, so in principle they could be helpful in the KiDS-Bright sample selection and calibration. However, their redshifts have lower than a 90\% probability of being correct \citep{GAMA-II}, and they are not recommended for scientific applications.} $\mathtt{NQ}\geq3$ and limited the redshifts to $z>0.002$ to avoid residual contamination by stars or local peculiar velocities. Thus cross-matching selected GAMA with KV imaging data yields
over 189\,000 sources with a mean redshift of $\meanz = 0.23$. When unambiguous, by "GAMA," from now on we mean this selection of GAMA galaxies in the equatorial fields.

A small fraction ($\sim4500$ in total) of GAMA galaxies do not have counterparts in the KiDS multiband catalog. About 1300 of these are located at the edges of the GAMA fields, where KV coverage did not reach. The rest are scattered around the equatorial fields and include a considerable fraction of $z<0.1$ galaxies, of low surface brightness galaxies, and of GAMA filler targets. These missing objects should not affect the analysis presented in this paper.

In Sect.~\ref{Sec:stellar_masses} we use the stellar mass estimates of GAMA galaxies for a comparison with our results from the KiDS-Bright catalog. For this we employ the StellarMassesLambdarv20 dataset, which includes physical parameters based on stellar population fits to rest-frame $u$-$Y$ SEDs, using Lambda Adaptive Multi-Band Deblending Algorithm in R \citep[LAMBDAR,][]{LAMBDAR-software} matched aperture photometry measurements of SDSS and VIKING photometry \citep{GAMA-LAMBDAR} for all $z < 0.65$ galaxies in the GAMA-II equatorial survey regions. This sample contains over 192\,000 galaxies, with a median $\log(M_\star/M_\sun) \sim 10.6$ assuming $\rm H_0 = 70 \,  km \, s^{-1} \, Mpc^{-1}$, and a range between the 1st and 99th percentile of $(8.4; 11.2)$ in the same units. Here and below by "log," we mean the decimal logarithm, $\log_{10}$.
For further details on the GAMA stellar mass derivation, see \cite{Taylor11} and \cite{GAMA-LAMBDAR}.

\subsection{KiDS-Bright galaxy sample}
\label{Sec:KiDS-bright-selection}

To ensure that the highly complete, flux-limited GAMA catalog is the appropriate \phz\ training set for the KiDS-Bright sample, the selection of the latter should mimic that of the former as closely as possible. The differences 
between the KiDS and SDSS photometry, filter transmission curves, as well as the data processing of both surveys prevent an exact matching. In particular, Petrosian magnitudes were not measured by the KiDS pipeline; even if they had been though, the different $r$-band PSF (subarcsecond in KiDS versus median $\sim1.3"$ in SDSS,) and depth ($\sim25$ mag of KiDS versus $\sim22.7$ in SDSS) would mean that the sources in common would on average have a much higher signal-to-noise in KiDS. Due to the photometric noise (Eddington bias, etc.), even applying the same cut to the same magnitude type (if possible) would not result in the same selection for the two surveys. 

Instead, we used the overlap with GAMA and designed an effective bright galaxy selection from KiDS, aiming at a trade-off between completeness and purity of the dataset. To select only extended sources (galaxies), we verified how the three star-galaxy separation metrics available in KiDS DR4 (\ttt{CLASS\_STAR}, \ttt{SG2DPHOT} and \ttt{SG\_FLAG}) perform for the GAMA sources. We found that the optimal approach is to jointly apply the following conditions: $\mtt{CLASS\_STAR}<0.5$ \& $\mtt{SG2DPHOT}=0$ \& $\mtt{SG\_FLAG}=1$. These remove less than $0.5\%$ of the matched KiDS$\times$GAMA $r_\mrm{Petro}<19.8$ galaxies, so this selection ensures a completeness of more than $\sim99.5\%$.

As far as the magnitude limit of the KiDS-Bright galaxy selection is concerned, we verified which of the $r$-band magnitude types -- \ttt{AUTO} or \ttt{ISO} -- is the most appropriate for the selection. 
%
%
We find that \ttt{ISO} matches the SDSS Petrosian magnitude slightly better: the median difference $\Delta_\mrm{iso}\equiv  r^\mrm{KiDS}_\mrm{iso} -  r^\mrm{GAMA}_\mrm{Petro} \simeq -0.02$ as compared to $\Delta_\mrm{auto}\simeq -0.06$. However, the scatter in $\Delta_\mrm{auto}$ is smaller than in $\Delta_\mrm{iso}$: the former is more peaked (i.e., narrower interquartile and 10- to 90-percentile ranges around the median) than the latter. We therefore decided to use $r_\mrm{auto}<20$ for the bright sample selection.
This ensures a completeness level of over $99\%$ with respect to the GAMA $r<19.8$ selection. 

\begin{figure}
\centering
\includegraphics[width=0.4\textwidth]{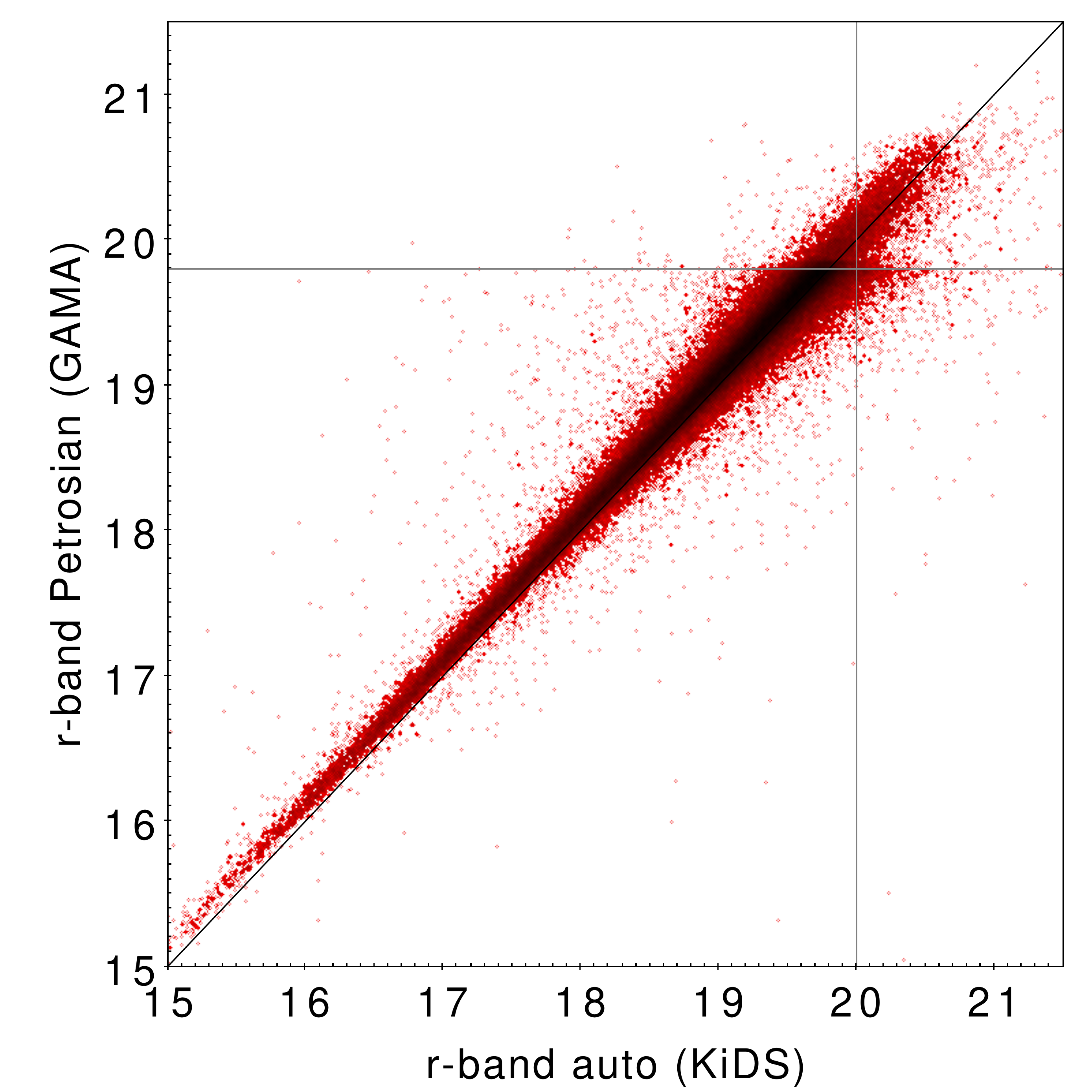} 
\caption{Comparison between the KiDS $r_\mrm{auto}$ and the Petrosian $r$-band from SDSS for galaxies in common between the two data sets.
The GAMA selection is based on the latter magnitude, whereas we
used the former to determine the flux limit of our galaxy sample. The relevant magnitude limits are indicated with the gray lines, and the black diagonal is the identity line. } 
\label{Fig:rPetro-rAuto} 
\end{figure}

Figure \ref{Fig:rPetro-rAuto} presents a comparison of the SDSS Petrosian and KiDS \ttt{AUTO} $r$-band magnitudes for the galaxies in common with GAMA, including those beyond the completeness limit of the latter. The vertical and horizontal gray lines show the GAMA flux limit and the cut we adopted for the selection of the KiDS-Bright galaxy sample,  respectively. The combination of $r_\mrm{auto}<20$ and the star removal results in an
incompleteness in the galaxy selection of $\sim1.2\%$ with respect to GAMA.

Quantifying the purity of the resulting KiDS-Bright sample is more challenging as this formally requires a complete flux-limited sample of spectroscopically confirmed galaxies, quasars, and stars deeper than GAMA. As such a dataset is not available at present, we assess the purity using indirect methods instead. Possible contaminants are artifacts, incorrectly classified stars, or quasars for which galaxy \phzs\ may be inaccurate (especially if at high-$z$).

A small fraction of the bright sources have nonphysical or otherwise spurious \phzs\ (derived as described in Sect.~\ref{Sec:photo-zs}), that is $\zph<0$ or $\zph > 1$; these constitute only $\sim 0.05\%$ of the sample after applying the default mask. The stellar contamination should be minimal as we combined three flags for galaxy selection, which should yield a robust classification for objects detected with a high signal-to-noise ratio. Indeed, a cross-match with the SDSS DR14 spectroscopic star sample \citep{SDSS.DR14} yields only 170 matches out of $\sim50\,000$ SDSS stars in the KiDS-North area; extrapolated to KiDS-South, this would imply a contamination of this type of at most 0.05\%. Although SDSS stars do not constitute a uniform and flux limited sample at this depth, this still supports our expectation that the star contamination should be negligible. We also do not expect quasars to be significant and problematic contaminants: a similar cross-match, but with SDSS DR14 spectroscopic quasars, results in about 650 common sources, of which 90\% have $\zsp<0.5$. 
Matching the KiDS-Bright data with a much more complete, photometrically selected sample of KiDS quasars derived by \cite{Nakoneczny20}, which covers the whole DR4 footprint, gives $\sim1400$ common objects, of which 90\% have $\smash{\zph^\mrm{QSO}<0.66}$ (the "QSO" superscript refers to the quasar \phz\ as derived in that work). Both of these tests suggest that the possible contamination with high-$z$ quasars is also a fraction of a percent. The \phzs\ of such residual quasars are worse than for the general galaxy sample, but the very small number of them does not influence the overall statistics and the quality of the dataset.

Finally we examine the impact of KiDS-Bright objects that are fainter than the completeness limit of GAMA, that is they have $r_\mrm{Petro}>19.8$ (see Fig.~\ref{Fig:rPetro-rAuto}). Following the analysis above, these are most likely galaxies, and as such should not be considered contaminants, but they are not well represented by the GAMA spectroscopic sample or, alternatively, not represented at all. The \phz\ estimates of such galaxies could be affected by the fact that their calibration is based on the incomplete and nonuniform sampling of GAMA filler targets beyond the nominal flux limit of the survey. 
On the other hand, the KiDS-Bright objects beyond the GAMA limit, but with colors similar to those included in the flux-limited spectroscopic sample, should still attain reliable \phzs.

One way to estimate the number of such faint-end sources is to compare the catalogs for the GAMA equatorial fields. After all the selections, the KiDS-Bright sample comprises less than 192\,000 galaxies, whereas the GAMA sample, with $r_\mrm{Petro}<19.8$, contains more than
182\,000 objects. The difference of approximately 9000 objects provides an upper limit of $\sim4.7\%$ for galaxies that are not fully represented in the GAMA catalog. The true fraction is likely below this number because only galaxies with misestimated \phzs\ based on extrapolation beyond GAMA should be considered as potentially problematic. Their number is difficult to estimate without a comparison against a complete flux-limited galaxy spectroscopic sample, deeper than GAMA and overlapping with KiDS. Such a dataset is presently unavailable; 
we can, however, estimate how many of the KiDS-Bright galaxies are similar to GAMA "filler" targets. In the cross-matched KiDS$\times$GAMA sample, there are about 4800 GAMA fillers with $r_\mrm{Petro}>19.8$ out of the $\sim146$k selected in the same way as the KiDS-Bright ($r_\mrm{auto}<20$ plus the galaxy selections and masking detailed above); this yields about 3.3\%. The \phz\ performance of such a filler sample is worse, but not catastrophic: their $\meandz \simeq 1.6\times10^{-3}$ and $\sigma_z \simeq 0.024(1+z)$ at a mean redshift of $\meanz = 0.33$. For those KiDS-Bright galaxies which are not represented in GAMA at all, we cannot reliably estimate the overall \phz\ performance: deeper spectroscopic samples overlapping with KiDS are not sufficiently complete.

To summarize, we estimate that the KiDS-Bright sample has a very high purity level close to 100\%, as contamination from stars, high-redshift quasars, or artifacts is at a small fraction of a percent.
There is, however, an inevitable mismatch with GAMA flux-limited selection, with up to 3\% of the galaxies in KiDS-Bright not being fully represented by GAMA spectroscopy. These could potentially have \phzs\ based on ML extrapolation that are less reliable.

\section{Photometric redshifts}
\label{Sec:photo-zs}

To obtain \phz\ estimates that are optimized for our sample of bright low-redshift galaxies, we took advantage of the large amount of 
spectroscopic calibration data. To do so, we used supervised ML in which a computer model (based on ANNs in our case) learns to map the input space of "features" (magnitudes) to the output (redshift) based on training examples, which in our case are the KiDS galaxies with a GAMA \spz. The trained model was subsequently applied to the entire "inference" dataset, which in our case is the galaxy sample selected as described in Sect.~\ref{Sec:KiDS-bright-selection}.

\begin{figure*}
\centering
\includegraphics[width=0.3\textwidth]{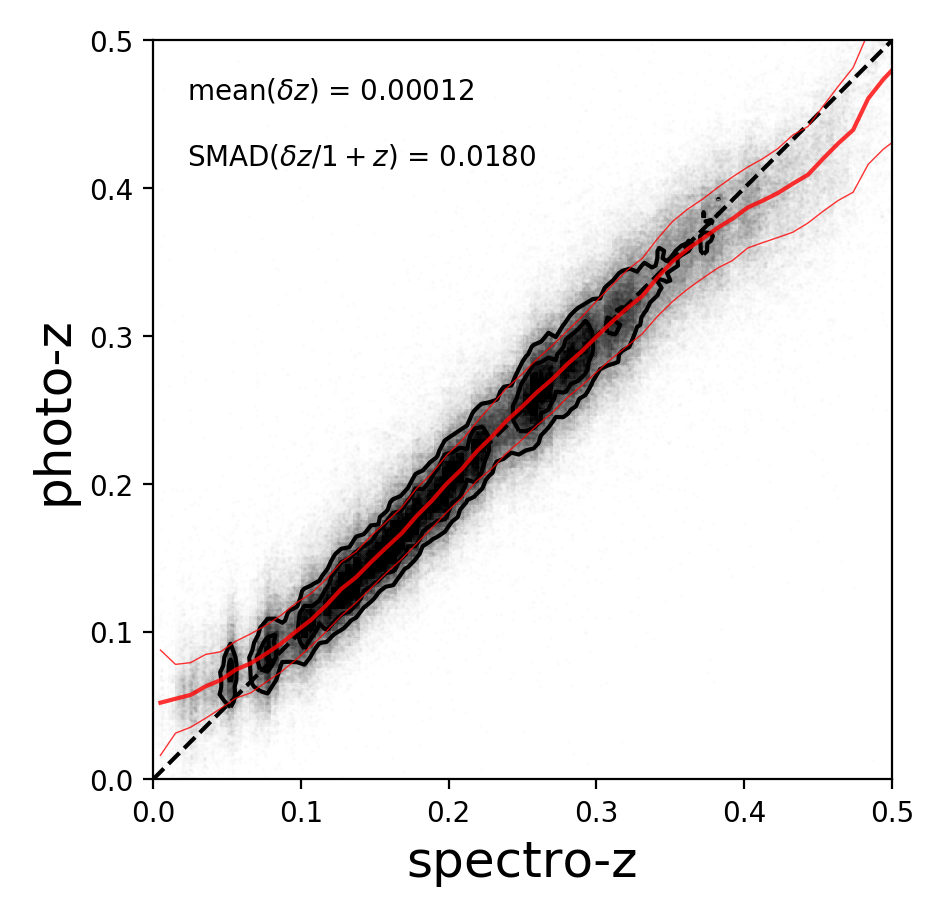}
\includegraphics[width=0.55\textwidth]{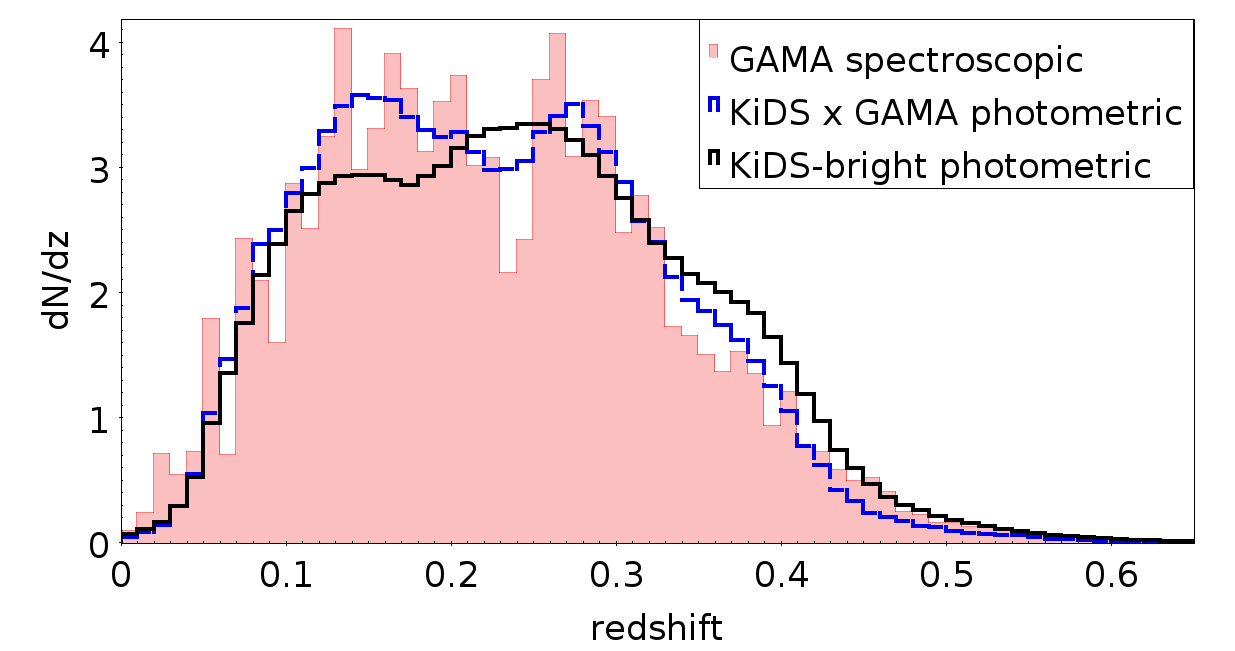}
\caption{Comparison of the KiDS-Bright photometric redshifts with the overlapping GAMA spectroscopic data. \textit{Left:} Direct \spz\ -- \phz\ comparison. The thick red line is the running median of the function $\zph(\zsp)$ and the thin red lines illustrate the scatter (SMAD) around the median. The black dashed line shows the identity. \textit{Right:} Comparison of redshift distributions of the GAMA spectroscopic training set (red bars), \phzs\ for the common KiDS$\times$GAMA sources (blue dashed line), and the full KiDS-Bright \phz\ sample (black line). The histograms are normalized to the unit area.}
\label{Fig:specz-photoz} 
\end{figure*}

Similarly to \citetalias{Bilicki18}, we used the ANNz2 software\footnote{Available for download from \url{https://github.com/IftachSadeh/ANNZ}. We used version \#2.3.1.} \citep{ANNz2} to derive the \phzs\ for the  KiDS-Bright galaxy sample. This package implements a number of supervised ML models for regression and classification. Throughout this work, we employed ANNz2 in the "randomized regression" mode, in which a preset number (here: 100) of networks with randomized configurations is generated for each training, and a weighted average is provided as the output. 
We trained ANNs using the GAMA-II equatorial sources that overlap with KiDS DR4. We verified that adding the Southern GAMA G23 data does not improve the final \phz\ statistics -- G23 is shallower and less complete than the equatorial data, and including it does not add any new information in the feature space that the networks could use to improve the \phz\ performance. For similar reasons, we did not employ other wide-angle spectroscopic data, such as SDSS or 2dFLenS, to the training set.
Those samples include flux-limited subsets shallower and less complete than GAMA, while at the fainter end they encompass only color-selected galaxies, mostly red ones, which if employed in \phz\ training, would bias the estimates against blue sources. 

The galaxies were used in various configurations for the \phz\ training, validation, and tests. To enable some level of extrapolation by the ML model in the range of $r_\mrm{Petro}>19.8$ \& $r_\mrm{auto}<20$ (see Fig. \ref{Fig:rPetro-rAuto}), we did not limit them to the GAMA completeness cut. As the ANNs in our setup cannot handle missing data, we require photometry in all nine bands, also for the tests we discuss below. However, as the galaxies are much brighter than the magnitude limits of both KiDS and VIKING, we only lose $\sim1500$ objects out of a total of 189\,000 spectroscopic galaxies. 

In the testing phase, we randomly selected 33\% of the galaxies with redshifts from GAMA as a joint training and validation set, while the rest were used for testing. In all cases, the actual validation set (used internally by ANNz2 for network optimization) was randomly selected as half of the input training and validation sample. For the final training of the \phzs\ of our bright galaxies, we used the entire cross-matched sample, again with a random half-half split for actual training and validation (optimization) in ANNz2. As shown in \citetalias{Bilicki18}, these proportions between training, validation, and test sets can be varied within reasonable ranges without much influence on the results; we dealt with sufficiently large samples to ensure robust statistics.

To evaluate the performance, we measured the "scatter," defined as the scaled median absolute deviation (SMAD) of the quantity $\Delta z \equiv  \delta_z/(1+z_\mrm{true})$ with $\delta_z \equiv \zph - z_\mrm{true}$ and $\mrm{SMAD}(x) = 1.4826  \times \mrm{median}(|x - \mrm{median}(x)|)$. As $z_\mrm{true}$, we used the \spzs\ from the test sample. In \citetalias{Bilicki18} we showed that adding NIR VIKING magnitudes to the $ugri$-only setup available in KiDS DR3 reduced the scatter of the \phzs\ at the GAMA depth by roughly 9\%, from $\sigma_z \simeq 0.022(1+z)$ to  $0.020(1+z)$. The VIKING measurements employed there were based on GAMA-LAMBDAR forced photometry \citep{GAMA-LAMBDAR}, using SDSS apertures as input and without PSF corrections that are applied in KV processing \citep{KV450-data,KiDS-DR4}. We therefore expect that the improved color measurements in DR4 should reduce the errors even further. 
Indeed, we find that the scatter of 9-band KiDS DR4 \phzs\ for our bright galaxies is further reduced  with respect to the KiDS DR3 + LAMBDAR VIKING statistics, in total by $\sim 18\%$ from the DR3 $ugri$-only derivation (see Table \ref{Tab: photo-z stats} below). We also verified that omitting any of the 9 bands worsens the performance. None of the VIKING bands stand out, which is expected, because  for the redshifts covered by GAMA ($z<0.5$), the NIR data do not trace clear features in the spectrum; rather they sample the Rayleigh-Jeans tail,
and thus each of the VIKING bands adds a similar amount of information.

\begin{figure*}
\centering
\includegraphics[width=0.47\textwidth]{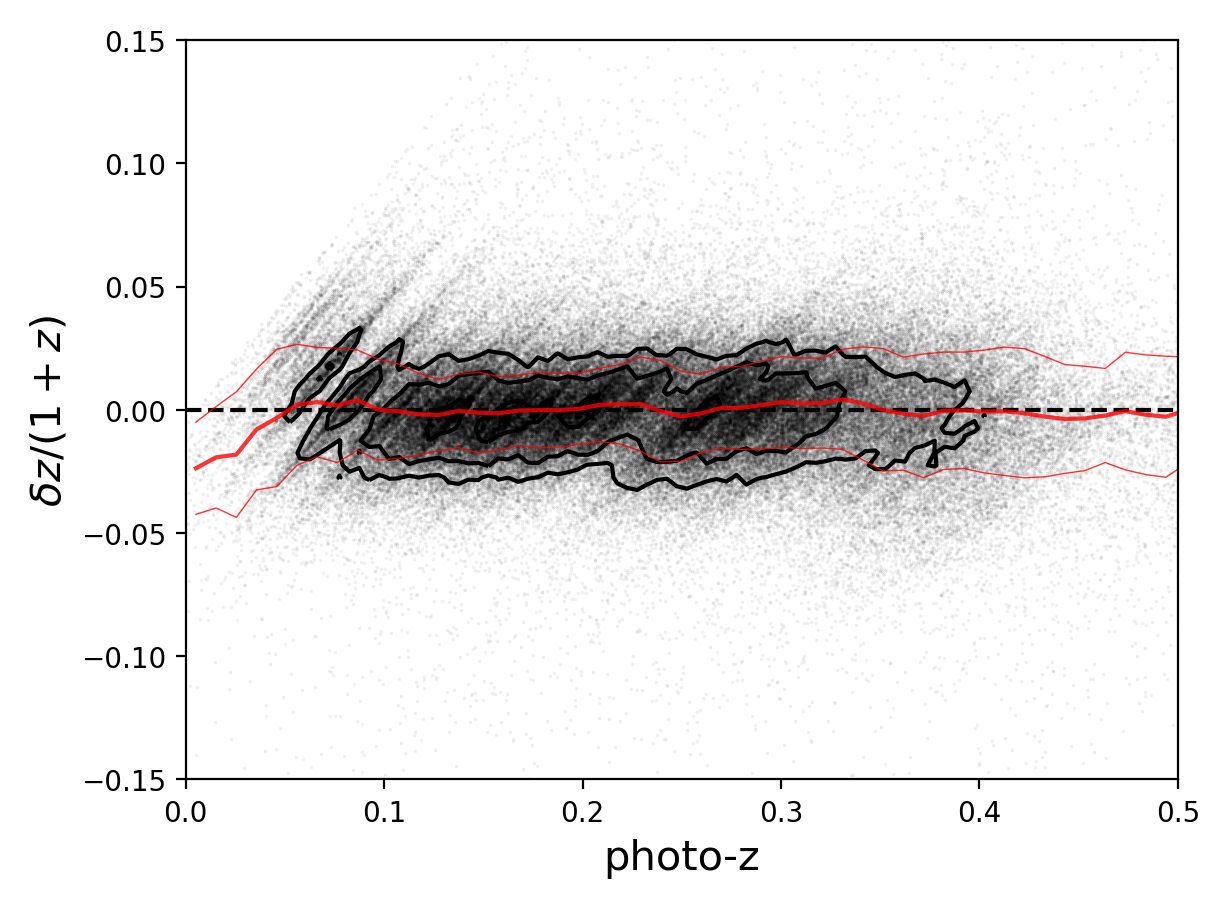}
\includegraphics[width=0.45\textwidth]{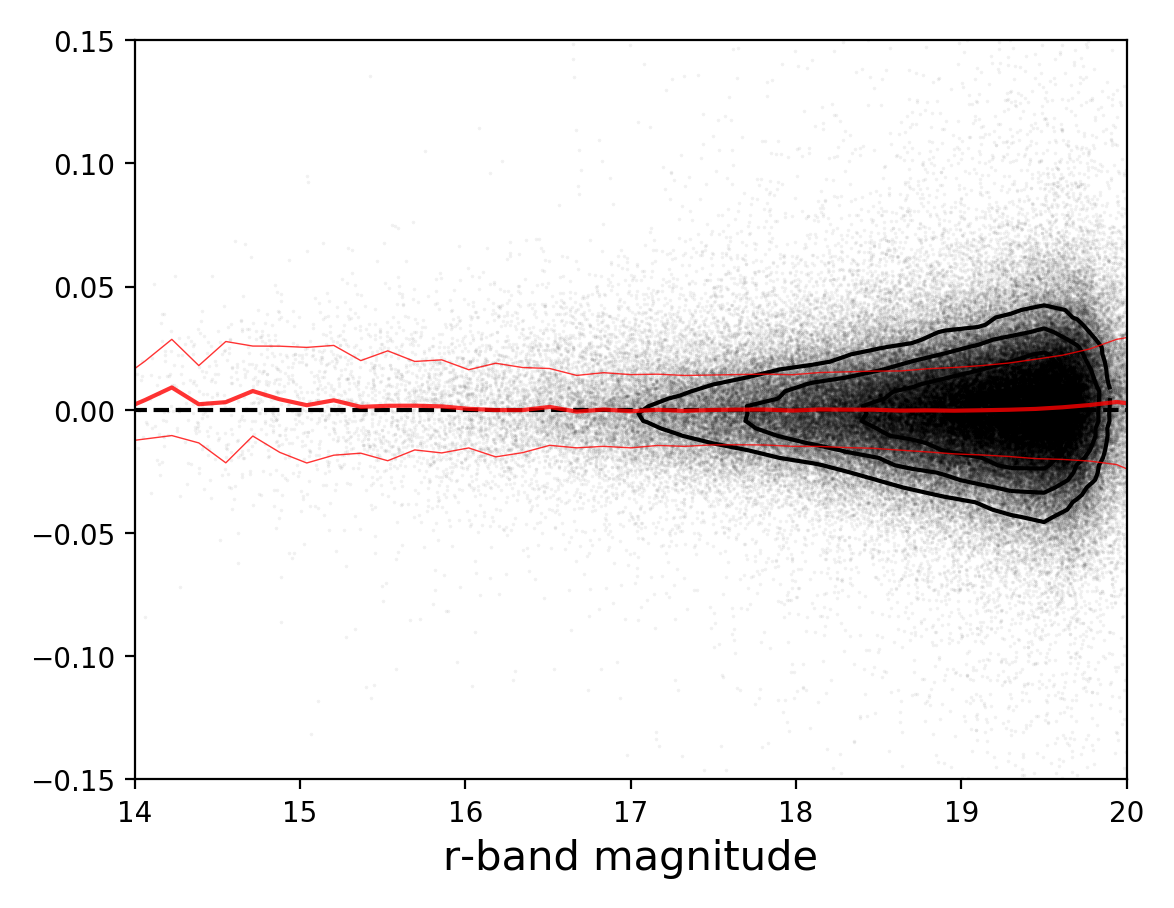}
\caption{Photometric redshift errors in the KiDS-Bright sample as a function of \phz\ \textit{(left)} and of the KiDS $r$-band \ttt{AUTO} magnitude \textit{(right)}, calibrated on overlapping GAMA data. Each dot is a galaxy, with contours overplotted in the highest number density areas. The thick red line is the running median and the thin red lines illustrate the scatter (SMAD) around the median. The stripes in the left panel originate from the large-scale structures present in the GAMA fields.}
\label{Fig:photoz-performance} 
\end{figure*}

The \phzs\ could be potentially improved if additional features are included in the training. In \citetalias{Bilicki18} we studied this in detail for a similar bright sample of galaxies, and found that adding colors (magnitude differences) and galaxy angular sizes (semi-axes of best-fit ellipses) did lead to better \phz\ estimates, compared to the magnitude-only case. For the 9-band data, however, there are 36 possible colors and feeding the ANNs with all of them, together with the magnitudes, would be very inefficient without specific network optimization each time; some prior feature importance quantification to choose the most relevant subset would be needed. This is beyond the scope of this work and therefore we limit the \phz\ derivation to magnitudes only. Unlike \citetalias{Bilicki18}, we decided not to use any size information
because the available estimates are not PSF-corrected. Using the uncorrected sizes could introduce a systematic variation of \phz\ quality with the PSF at a source position. As one of the applications of the KiDS-Bright sample is to use it for cosmological measurements, we decided to employ only the PSF-corrected \textsc{GAaP} magnitudes for redshift estimation.

As already mentioned in Sect.~\ref{Sec:KiDS_data}, each KiDS object is assigned a \ttt{MASK} flag, indicating issues with source extraction. The default masking, used to create the KiDS-1000 weak lensing mosaic catalogs, is to remove the sources matching bit-wise the value 28668. We checked the importance of this masking for \phz\ quality by performing two ANN trainings: one including all the training sources with any mask flag, and another one where only the sources with the default masking were used. For each of the cases, the performance was evaluated using the same blind test set. We did not observe any difference between the \phz\ statistics for the two training cases. Our interpretation is that the ANNs are able to "learn" the noise related to the \ttt{MASK} flag. By ignoring this flag in the training phase, they are still able to robustly estimate \phzs. At the same time, as far as the evaluation is concerned, there is a clear deterioration in the \phz\ performance for the sources that should be masked out with respect to those that pass the default selection, for both training setups. Motivated by these findings, we ignored the \ttt{MASK} value for the training set for the final sample. We however provide a flag with our \phz\ estimates that indicates which of the galaxies meet the condition $(\mtt{MASK}\; \&\; 28668)>0$ and should  be preferably masked out for science applications.

\subsection{Photometric redshift performance}
\label{sec:photo-z-performance}

We compare the KiDS-Bright \phzs\ with the overlapping \spzs\ from GAMA in Fig.~\ref{Fig:specz-photoz}. The left panel shows that the \phzs\ are overestimated at low-$z$ and underestimated at high-$z$,
which is common for ML approaches. Nonetheless, the overall performance is excellent, with a low average bias and a small and near constant scatter as a function of redshift. 

The redshift distributions presented in the right panel of Fig.~\ref{Fig:specz-photoz} indicate that for the matched KiDS$\times$GAMA galaxies, $dN/d\zph$ (blue dashed line) closely follows the general shape of the true $dN/d\zsp$ (red bars), preserving even the "dip" observed in GAMA at $z\sim0.25$ (emerging by chance due to large-scale structures passing through the equatorial fields; e.g., \citealt{Eardley15}). As far as the redshift distribution of the full photometric sample is concerned (black solid line), we observe some piling up of \phzs\ at the very same range where the GAMA dip is present, but also at $\zph \sim 0.35$. This might be the result of the extrapolation by ANNz2 in the regime $r_\mrm{auto} \sim 20$, where sources can be fainter than the GAMA completeness limit (Fig.~\ref{Fig:rPetro-rAuto}), or for sources that are for some other reason under-represented in GAMA (as discussed 
in Sect.~\ref{Sec:KiDS-bright-selection}). 

\begin{table*}
\begin{center}
\caption{Statistics of photometric redshift performance for the KiDS-Bright sample and selected subsamples. The sample sizes refer to the full photometric selection.\label{Tab: photo-z stats}}
\begin{tabular}{l r r r r r r }
\hline\hline
\centering { \textbf{sample} } & { \textbf{number of} } & \multicolumn{1}{c}{ \textbf{mean} } & \multicolumn{1}{c}{ \textbf{mean of} } & { \textbf{mean of} } & \multicolumn{1}{c}{\textbf{st.dev.\ of}} & \multicolumn{1}{c}{\textbf{SMAD of}} \\
{} & { \textbf{galaxies} } & \multicolumn{1}{c}{ \textbf{redshift} } & \multicolumn{1}{c}{$\delta z = z_\mathrm{ph}-z_\mathrm{sp}$ } &  \multicolumn{1}{c}{$\delta z/(1+z_\mathrm{sp})$ } &  \multicolumn{1}{c}{$\delta z/(1+z_\mathrm{sp})$ } &  \multicolumn{1}{c}{$\delta z/(1+z_\mathrm{sp})$} \\ 
\hline
full KiDS-Bright \tablefootmark{a} & $1.24\times10^6$ & 0.226 & $1.2\times10^{-4} $ & $6.7\times10^{-4} $ & 0.0246 & 0.0180  \\
after masking \tablefootmark{b} & $1.00\times10^6$ & 0.229 & $4.6\times10^{-4} $ & $9.0\times10^{-4} $ & 0.0237 & 0.0178  \\
red galaxies \tablefootmark{c} & $3.91\times10^5$ & 0.243 & $-2.7\times10^{-4} $ & $2.0\times10^{-4} $ & 0.0194 & 0.0159  \\
blue galaxies \tablefootmark{c} & $4.25\times10^5$ & 0.212 & $1.5\times10^{-3} $ & $1.8\times10^{-3} $ & 0.0274 & 0.0200  \\
luminous red galaxies \tablefootmark{d} & $7.18\times10^4$ & 0.305 & $1.1\times10^{-3} $ & $1.1\times10^{-3} $ & 0.0161 & 0.0141  \\
\hline 
\end{tabular}
\end{center}
\begin{small}
\tablefoot{}
\tablefoottext{a}{Flux-limited galaxy sample ($r_\mrm{AUTO}<20$); see Sect.~\ref{Sec:KiDS-bright-selection} for other details of the selection.\\}
\tablefoottext{b}{Using the KiDS \ttt{MASK} flag, removing the sources meeting the condition $(\mtt{MASK} \& 28668)>0$ (bit-wise).\\}
\tablefoottext{c}{Selected using the $r$-band absolute magnitude and rest-frame $u-g$ color based on \textsc{LePhare} output; see Sect.~\ref{Sec:stellar_masses} for details.\\}
\tablefoottext{d}{Selected using the Bayesian model detailed in \cite{Vakili20},  jointly encompassing the "dense" and "luminous" samples. Numbers refer to the LRGs overlapping with the KiDS-Bright sample and the \phz\ statistics are based on the ANNz2 derivations.}
\end{small}
\end{table*}

To illustrate the KiDS-Bright \phz\ performance in more detail,
we show the redshift errors $\delta z/(1+z)$ as a function of
\phz\ and $r$-band magnitude in Fig.~\ref{Fig:photoz-performance}. The errors show little dependence on the $r$-band magnitude or photometric redshift, except for the range $\zph < 0.05$. Since at this redshift range  the number density of the photometric KiDS galaxies is very small, and as it is additionally very well covered by wide-angle spectroscopic samples such as SDSS Main \citep{SDSS.MGS}, 6dFGS \citep{6dF.DR3}, and GAMA itself, this worse \phz\ performance is irrelevant for scientific applications of the KiDS-Bright sample. We however recommend using only the $\zph>0.05$ sources; this cut affects less than 1\% of the sample. At the high-redshift end of the dataset, $\zph \gtrsim 0.4$, both the KiDS-Bright and GAMA calibration samples become very sparse (Fig.~\ref{Fig:specz-photoz}). However, the \phz\ quality remains comparable to the rest of the dataset (Fig.~\ref{Fig:photoz-performance}), so the galaxies with $\zph \lesssim 0.5$ should be safe for scientific applications once the flux-limited character of the sample is taken into account.

The fact that the \phzs\ are practically unbiased as a function of the \phz\ is typical for ML-based derivations, as the algorithms usually search for the best-fit solution for the \phzs\ as calibrated on known \spzs. As the redshift distribution is not constant but has a well-defined maximum (especially for flux-limited samples such as ours), this then leads to an inevitable bias as a function of \spz\ at the extremes of the coverage, as already illustrated in Fig.~\ref{Fig:specz-photoz}. 
However, in most applications it is important to be able to select in \phz\ and calibrate the true redshift distribution of a given sample a posteriori (e.g., in \phz\ bins). For this, knowledge of the \phz\ error distribution (discussed below in Sect.~\ref{Sec:photo-z-model}) plus the $dN/d\zph$ are usually sufficient to build a reliable model.

The relative paucity of $\zsp\sim0.25$ galaxies in the GAMA-equatorial data, used here for the \phz\ training, is caused by a large-scale structure in these fields. This could potentially affect our redshift estimates if it was spuriously propagated by ANNz2. As we have already pointed out, this dip is correctly reproduced in the $dN/d\zph$ of the matched GAMA$\times$KiDS sample, but it is not present in the overall \phz\ distribution of the full KiDS-Bright sample. This suggests that the training is not significantly affected.
As an additional test, we compared $dN/d\zsp$ and $dN/d\zph$ of a cross-match between KiDS-Bright and spectroscopic redshifts in the Southern GAMA G23 field, in which such a lack of $z\sim0.25$ sources is not observed. As mentioned earlier, the latter dataset was not used for the \phz\ training because it is shallower and less complete than the GAMA-equatorial data. A comparison of the redshift histograms shows no  spurious lack of \phzs\ at $z\sim0.25$. Nonetheless, close inspection of the left-hand panel of Fig.~\ref{Fig:photoz-performance} does suggest some variation in \phz\ performance in this range; a similar effect is also observed in a $\zsp$ versus $\delta z$ comparison. Such "wiggles" in the \phz\ error as a function of redshift are still present if the G23 data are added to the ANNz2 training.
However, for the current and planned applications of the KiDS-Bright sample, these issues are not significant. Nonetheless, this might need revisiting for future analyses with the full-area KiDS DR5 data.

Table \ref{Tab: photo-z stats} provides basic  \phz\ statistics for our KiDS-Bright sample. We list the total number of sources, their mean redshift, as well as \phz\ bias and scatter (evaluated on overlapping GAMA spectroscopy). Comparison of the statistics for the full KiDS-Bright sample with that after masking demonstrates that masking improves the \phz\ statistics somewhat; interestingly, it also slightly enlarges the mean redshift. We also report results when the sample is split by color based on the $r$-band absolute magnitude and the rest-frame $u-g$ color, derived with \textsc{LePhare}, as detailed in Sect.~\ref{Sec:stellar_masses}. With the adopted split, the red galaxies are slightly less numerous than the blue ones, but their \phz\ performance is noticeably better. 

For reference we also provide the results for the galaxies that overlap with the LRG sample from \cite{Vakili20}, but using our ANNz2 redshift estimates. This particular subsample stands out with $\mrm{SMAD}(\Delta z) \sim 0.014$, albeit with a slightly larger overall bias  of $\meandz \sim 10^{-3}$,
which is still over an order of magnitude smaller than the scatter. These values are comparable to those obtained in \cite{Vakili20} using the dedicated red-sequence model, which confirms the excellent quality of our \phzs. The blue galaxies, despite performing worse overall in terms of their \phz\ statistics, still have very well constrained redshifts with $\mrm{SMAD}(\Delta z) \simeq 0.02$. 
For the blue and red galaxies, we find similar trends as the ones presented in  Fig.~\ref{Fig:photoz-performance} for the full sample, albeit with different levels of scatter.

The quality of \phzs\ can vary as a function of various survey properties. In Appendix \ref{App: Systematics} we present a short summary of the \phz\ error variation for the KiDS-Bright sample versus a number of both KiDS-internal (PSF, background, and limiting magnitudes) and external (star density and Galactic extinction) observational effects. We find that both the \phz\ bias and scatter are generally stable with respect to these quantities. 

As mentioned earlier, the \phz\ performance could potentially be improved if we additionally used galaxy colors, sizes, or other measurements correlated with redshift (e.g.,\ \citetalias{Bilicki18}). Some other avenues to explore here are to employ dedicated photometry, which is targeted better at apparently bright and large galaxies, such as those in our sample; this exists already for the GAMA fields \citep{Bellstedt20} and will be available for the entire KiDS footprint from the 4MOST WAVES team \citep{WAVES}. One could also inspect other empirical \phz\ methods, for instance the "scaled flux matching" \citep{Baldry21} or even deep learning techniques \citep[e.g.,][]{Pasquet19}. We plan to study these aspects in the final KiDS data, as well as in the mocks resembling the KiDS-Bright sample \citep{KiDS-mocks}, to verify how much improvement in the \phzs\ we can still hope for.

\subsection{Analytical model of the redshift errors}
\label{Sec:photo-z-model}

For a number of applications, such as angular clustering, GGL, or cross-correlations with other cosmological tracers, it is useful to have an analytical model of the \phz\ errors, which can be used in theoretical predictions \citep[e.g.,][]{BA18,PB18,Hang20}. The photometric redshift error distribution usually departs from a Gaussian shape due to a considerable number of several-$\sigma$ outliers and generally broader "wings" \citep[e.g.,][]{2MPZ,Pasquet19,Beck21}. This is why SMAD, or alternatively percentiles \citep[e.g.,][]{2dFLenS-photo-z,Soo18,Alarcon20}, are better suited to quantify the \phz\ scatter than the standard deviation, which is sensitive to outliers. Functional forms to fit the empirical \phz\ error distribution include the "modified Lorentzian" \citep{2MPZ,PB18,Hang20} or the Student's t-distribution \citep{Vakili20}. The former is provided by \citep{2MPZ}
\begin{equation} \label{Eq:gen-Lorentz}
 N(\Delta z) \propto \left(1+\frac{\Delta z^2}{2 a s^2}\right)^{-a},  
\end{equation}
where we have assumed that the \phzs\ are on average unbiased, which is a good approximation in our case as $\langle \Delta z \rangle \ll \mrm{SMAD}(\Delta z)$ (see Table \ref{Tab: photo-z stats}). This can be easily generalized to the case of non-negligible bias by introducing an extra parameter \citep{Hang20}. In Eqn.~\eqref{Eq:gen-Lorentz}, the parameter $s$ is related to the width of the distribution, while $a$ encodes the extent of the wings. We note that both $a$ and $s$ can be parameterized as \phz-dependent to build an analytical model of redshift error \citep{PB18}.

\begin{figure}
\centering
\includegraphics[width=0.5\textwidth]{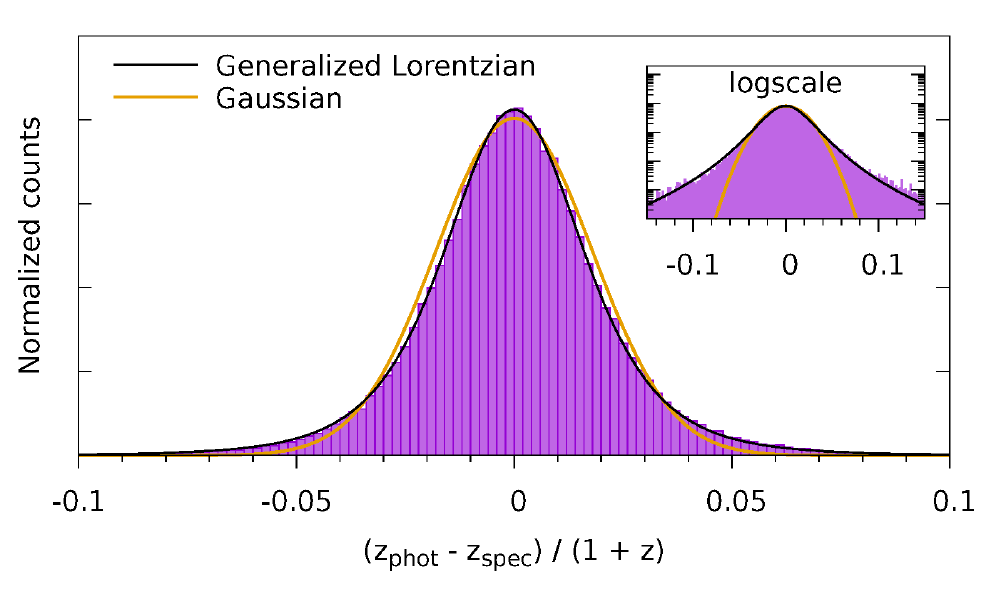}
\caption{Histogram of photometric redshift errors in the KiDS-Bright sample (magenta bars) fitted with a generalized Lorentzian (Eqn.\ \ref{Eq:gen-Lorentz}, black line) with parameters $a=2.613$ and $s=0.0149$, compared to best-fit Gaussian (orange) with $\sigma=0.0180$. The top-right inset elucidates the differences in the wings as seen in log-scaling.}
\label{Fig:photoz-errors} 
\end{figure}

We used  Eqn.~\eqref{Eq:gen-Lorentz} to fit the \phz\ error distribution in the KiDS-Bright sample and find the best-fit parameters to be $a=2.613$ and $s=0.0149$. Qualitatively, this is indeed a very good fit to the $\Delta z$ histogram, as illustrated in Fig.~\ref{Fig:photoz-errors}, clearly outperforming the best-fit Gaussian with $\sigma=0.0180$ (also assuming an average zero bias). The inset, with a log-scale to highlight the wings, shows that the Gaussian fails to account for the outliers.
We do not quantify the goodness of fit of the two models as we do not have meaningful information on the errors on the $\Delta z$ histogram.

Such an empirical shape for the \phz\ error distribution could be the result of combining galaxy populations with different redshift precision, as illustrated in Table \ref{Tab: photo-z stats} for the red and blue selections. We have however verified that for both these galaxy subpopulations, selected as described in Sect.~\ref{Sec:stellar_masses} below, their respective $\Delta z$ distributions are also better fitted with the modified Lorentzian \eqref{Eq:gen-Lorentz} than with a Gaussian. Similarly, as shown in \cite{Vakili20}, \phzs\ for luminous red galaxies, albeit derived with a different approach, also display a non-Gaussian error distribution. This might indicate that such an error distribution is an inevitable outcome of at least some empirical \phz\ methodologies.

\section{Stellar masses \& rest-frame absolute magnitudes}
\label{Sec:stellar_masses}

We estimate a number of rest-frame properties for each KiDS-Bright galaxy in the same manner as was done for the full-depth KV data within the DR3 footprint \citep[KV450,][]{KV450-data}. We did this by fitting model spectral energy distributions (SEDs) to the 9-band  \textsc{GAaP} fluxes of each galaxy using the \textsc{LePhare} \citep{Arnouts99, Ilbert06} template fitting code. In these fits, we employed our ANNz2 \phz\ estimates as input redshifts for each source, treating them as if they were exact. In practice, this has little influence on the fidelity of the stellar mass estimates (see \cite{Taylor11}). We used a standard concordance cosmology ($\Omega_{\rm m} = 0.3$, $\Omega_\Lambda = 0.7$, and $\rm H_0 = 70 \,  km \, s^{-1} \, Mpc^{-1}$), a \cite{Chabrier03} initial mass function, the \cite{Calzetti94} dust-extinction law, \cite{BC03} stellar population synthesis models, and exponentially declining star formation histories. The input photometry to \textsc{LePhare} was extinction corrected using the \cite{SFD} maps with the \cite{SF11} coefficients, as described in \cite{KiDS-DR4}. For the optical VST bands, we utilized the filter profiles measured at the center of the field of view, available from the ESO webpages\footnote{\url{https://www.eso.org/sci/facilities/paranal/instruments/omegacam/inst.html}}. 
For the NIR VISTA data, we used the averaged filter profile of all 16 filter segments per band \citep{VIKING}.

The \textsc{LePhare} code returns a number of quantities for each source, detailed in Appendix \ref{App:data_release}. The best-fit \ttt{MASS\_BEST} is the one that should be used as the estimate of galaxy's stellar mass; this quantity is available for almost all KiDS-Bright objects, except for a few hundred which have unreliable \phzs\ (e.g., $\zph<0$). When using these stellar mass estimates, it is however important to take into account the "flux scale correction" related to the fact that the \textsc{GAaP} magnitudes used by \textsc{LePhare} underestimate fluxes of large galaxies. The correction that we used is based on the difference between the AUTO and \textsc{GAaP} $r$-band magnitudes (see Eqn.~\ref{Eq:fluxscale}) and it was added to the logarithm of the stellar mass estimate given by \ttt{MASS\_BEST} (Eqn.~\ref{Eq:Mstar-tot}).

The code also outputs \ttt{MASS\_MED}, which is the median of the galaxy template stellar mass probability distribution function. This quantity can take a value of $-99$, which indicates that a  galaxy was best-fit by a non-galaxy template; although, the \ttt{MASS\_BEST} value still reports the mass from the  best-fitting galaxy template. In some cases this could highlight stellar contamination for sources that are best-fit by a stellar template and additionally have a small flux radius, and this could be even used for star-galaxy separation (see the related discussion in \citealt{KV450-data}). This is, however, not a concern for our sample: out of over 270\,000 objects with $\mtt{MASS\_MED}=-99$, only a few lie on the stellar locus. This further confirms the very high purity level of the KiDS-Bright catalog, as already concluded in Sect.~\ref{Sec:KiDS-bright-selection}.

\begin{figure}
\centering
        \includegraphics[width=0.9\columnwidth]{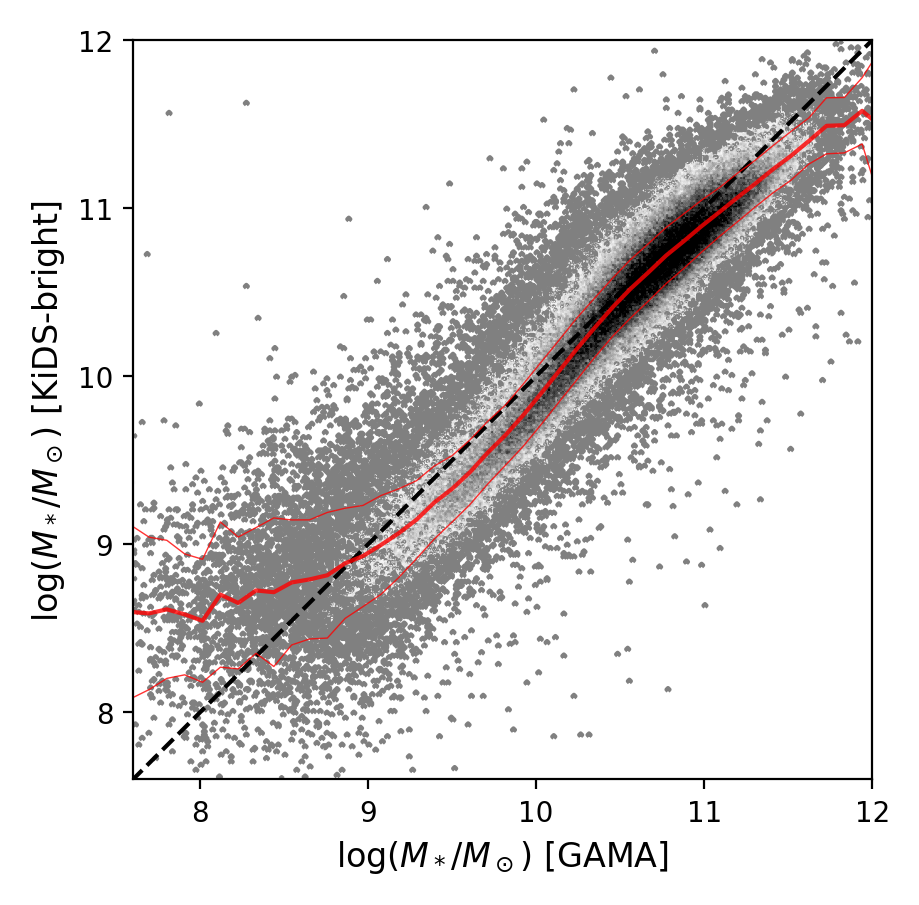}
        \caption{Comparison of the derived stellar masses between the photometric KiDS-Bright sample (this work) and the spectroscopic GAMA dataset for galaxies common to both catalogs. The light gray to black scaling illustrates the bulk of the sample, while the outliers where the number density is smaller, are shown with individual large gray dots. The thick red line is the running median, and thin red lines illustrate the scatter (SMAD).}      \label{fig:stellar-mass-compare}
\end{figure}

The median stellar mass of the KiDS-Bright sample is $\log(M_\star/M_\sun)\sim10.5$, with a range between the 1st and 99th percentile of roughly $8.5 < \log(M_\star/M_\sun) < 11.4$. In order to assess the quality of these stellar mass estimates, we compared them with the GAMA stellar mass catalog \citep{Taylor11,GAMA-LAMBDAR}, introduced in Sect.~\ref{Sec:GAMA}. First of all, it is worth noting that the overall distributions of the stellar masses (normalized histograms of $\mrm{d}N/\mrm{d}(\log M_\star)$) are very similar, and in particular their maximum (mode) is at $\sim10.75$ in both cases. Cross-matching the two samples gives about 145\,000 galaxies with stellar masses from both KiDS-Bright and GAMA. We compare these directly in Fig.~\ref{fig:stellar-mass-compare}, where we also plotted the running median relation together with the corresponding SMAD (respectively thick and thin red lines). We see that the relation is within $\sim1\sigma$ from the identity line (dashed) over a wide range in stellar mass and significantly departs from it  only at the tails of the distribution. On average, the KiDS-Bright stellar mass estimates are smaller than those of GAMA by $\Delta \log M_\star \equiv \log M^\mrm{KiDS}_* - \log M^\mrm{GAMA}_* = -0.09 \pm0.18$ dex (median and SMAD). Such an overall bias between the former and the latter is expected: while our flux-scale correction is meant to compensate for the flux missed by the \textsc{GAaP} measurements with respect to \ttt{AUTO} magnitudes, an analogous correction in GAMA serves to account for a flux that falls beyond the finite SDSS-based \ttt{AUTO} aperture used for the SEDs.

Nonetheless, the overall consistency is remarkable, given that the stellar masses were determined using different data and methodology: GAMA employed  spectroscopic redshifts together with LAMBDAR photometry from SDSS+VIKING $u$ to $Y$ bands, while we used \phzs\ and \textsc{GAaP} KiDS+VIKING $u$ to $K_\mrm{s}$ measurements. While the GAMA stellar masses cannot be treated as the "ground truth" due to inevitable systematics in the modeling, it is worthwhile to explore trends in the stellar mass differences between the two data sets. We observe no significant trend of $\Delta \log M_\star$ with magnitude or with color. Not surprisingly, the use of \phzs\ does affect the performance for galaxies, especially at very low redshifts ($\zsp\lesssim 0.07$). 

In general, we observe a linear trend in $\Delta \log M_\star$ with $\delta z/(1+z)$. If we account for this trend, the SMAD in $\Delta \log M_\star$ is $\sim 0.17$ dex, that is $\sim9\%$ lower than for the entire matched sample; this difference can be regarded as the effective increase in the scatter between GAMA and KiDS-Bright stellar mass derivations due to the \phzs\ only.  Overall, we find that the results are robust, with roughly constant scatter, if we select galaxies with $\zph>0.1$, for which the SMAD in $\Delta \log M_\star$ reduces to $\sim0.17$ dex. Therefore we restrict the GGL analysis presented in the next section to this redshift range; the removed galaxies would not be of much importance for the lensing analysis in any case.

\begin{figure}
        \includegraphics[width=\columnwidth]{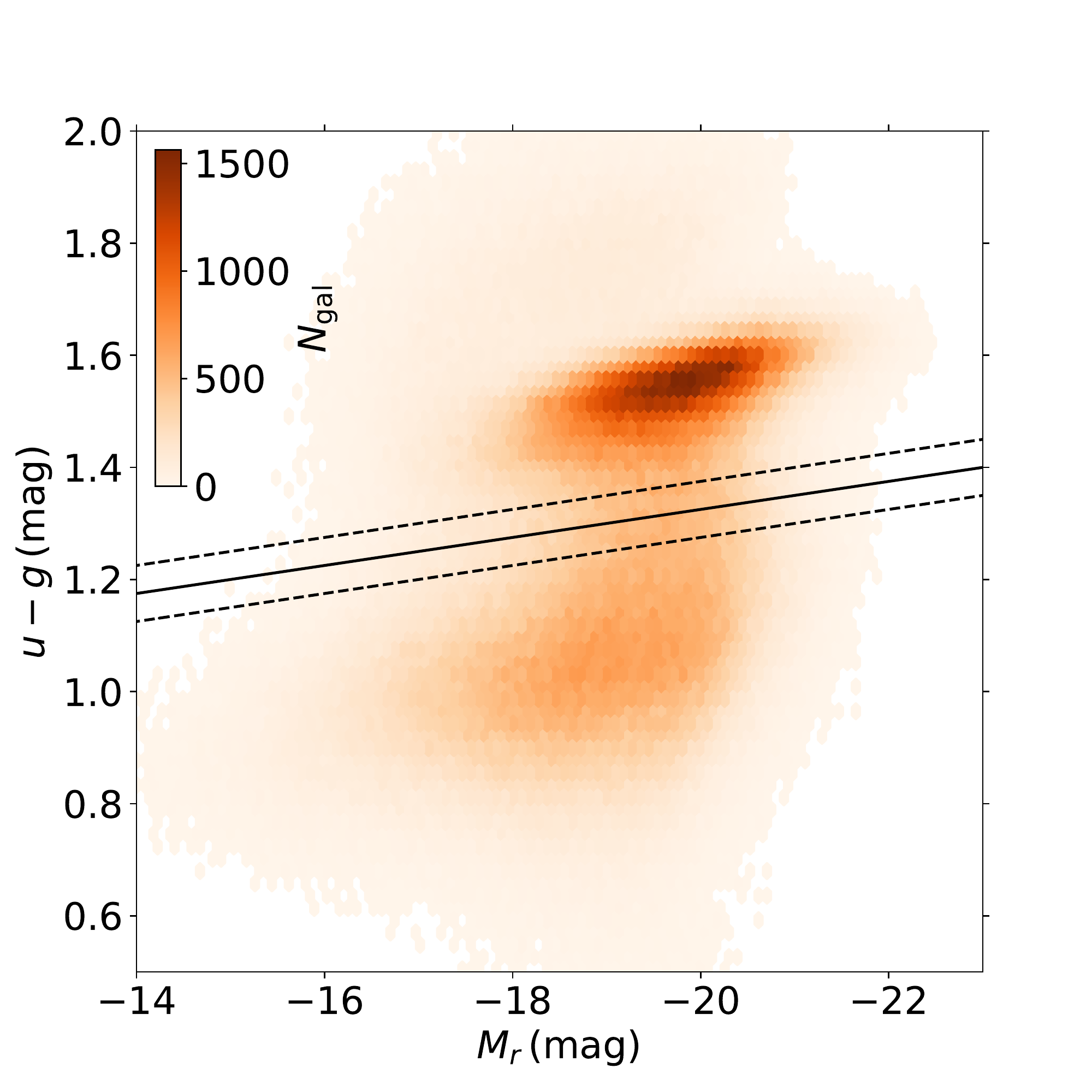}
        \caption{Distribution of the $u-g$ rest-frame color versus absolute $r$-band magnitude for the KiDS-Bright galaxy sample, based on \textsc{Le\-Phare} derivations with ANNz2 \phzs\ as input redshifts. We used the location of the green valley to derive an empirical split into red and blue galaxies above the upper dashed and below the lower dashed lines, respectively.}
        \label{fig:red-blue-split}
\end{figure}

We used the absolute $r$-band magnitude and the rest-frame $u-g$ color derived with \textsc{LePhare} (employing the ANNz2 \phzs\ as input redshifts) to select red and blue galaxies based on an empirical cut through the green valley in the color-magnitude diagram. We identified the ridge of the blue cloud to define the slope and locate the minimum at the absolute magnitude of $M_r = -19$. This results in a line that delimits the red and blue sample:
\begin{equation}
u-g = 0.825 - 0.025 \, M_r\,.
\end{equation}
Based on this cut, we define our red sample as those galaxies whose $u-g$ color is at least 0.05 mag above the cut line and the blue sample as those whose color is at least 0.05 mag below the line. The color-magnitude distribution and the cut through the green valley are shown in Fig.~\ref{fig:red-blue-split}. The \phz\ statistics for the red and blue galaxies defined this way have been presented in Sect.~\ref{Sec:photo-zs}; below in Sect.~\ref{Sec:GGL}, we use this split as well as the stellar masses in GGL measurements.

\section{Galaxy-galaxy lensing measurements}
\label{Sec:GGL}

As shown in the previous section, the excellent photometric redshift estimates for the galaxies in the KiDS-Bright sample allow for robust estimates of their physical characteristics, in particular the stellar mass. In this section we combine this information with accurate shape measurements for more distant KiDS sources from \cite{Giblin20} to measure the GGL signal. We first compared the lensing signal for a similar selection of lenses from GAMA and KiDS around the mode of the stellar mass distribution. We then split the sample of bright lens galaxies into blue and red subsamples (see Sect.~\ref{Sec:stellar_masses} and Fig.~\ref{fig:red-blue-split}), which were subsequently subdivided by stellar mass. To quantify the weak gravitational lensing signal, we used source galaxies from KiDS DR4 with a BPZ \phz\ in the range $0.1<z_\mrm{B}<1.2$. 

The lensing signal of an individual lens is too small to be detected, and hence we computed a weighted average of the tangential ellipticity $\epsilon_{\rm t}$ as a function of projected distance $r_{\rm p}$ using a large number of 
lens-source pairs. In the weak lensing regime, this provides an unbiased estimate of the tangential shear, $\gamma_{\rm t}$, which in turn can be related to the excess surface density (ESD) $\Delta\Sigma(r_{\rm p})$, defined as the difference between the mean projected surface mass density inside a projected radius $r_{\rm p}$ and the mean surface density at $r_{\rm p}$.

We computed a weighted average to account for the variation in the precision of the shear estimate, captured 
by the \emph{lens}fit weight $w_{\text{s}}$ \citep[see][for details]{FC17,Kannawadi19}, and the fact that the amplitude of the lensing signal depends on the source redshift. The weight assigned to each lens-source pair is
\begin{equation}
\label{eq:weights}
\widetilde{w}_{\mathrm{ls}}=w_{\mathrm{s}} \left(\widetilde \Sigma_{\mathrm{cr, ls}}^{-1}\right)^{2} \, ,
\end{equation}
the product of the \emph{lens}fit weight $w_{\text{s}}$ and the square of $\widetilde\Sigma_{\mathrm{cr, ls}}^{-1}$ -- the effective inverse critical surface mass density, which is a geometric term that downweights lens-source pairs that are close in redshift \citep[e.g.,][]{Bartelmann1999}.

We computed the effective inverse critical surface mass density for each lens using the  \phz\ of the lens $z_{\mathrm{l}}$ and the full normalized redshift probability density of the sources, $n(z_{\mathrm{s}})$. The latter was calculated by employing the self-organizing map calibration method, originally presented in \cite{KV450-SOM},  and then applied to KiDS DR4 in \citet{Hildebrandt20b}. The resulting effective inverse critical surface density can be written as follows:
\begin{equation}
\label{eq:crit_effective}
\widetilde\Sigma_{\mathrm{cr, ls}}^{-1}=\frac{4\pi G}{c^2} \int_{0}^{\infty} (1+ z_{\rm l})^{2}  D(z_{\mathrm{l}}) \left(\int_{z_{\mathrm{l}}}^{\infty} \frac{D(z_{\mathrm{l}},z_{\mathrm{s}})}{D(z_{\mathrm{s}})}n(z_{\mathrm{s}})\, \mathrm{d}z_{\mathrm{s}} \right)\, p(z_{\mathrm{l}})\, \mathrm{d}z_{\mathrm{l}} \, ,
\end{equation}
where $D(z_{\rm l})$, $D(z_{\rm s})$, and $D(z_{\rm l},z_{\rm s})$ are the angular diameter distances to the lens, source, and between the lens and the source, respectively.

For the lens redshifts $z_{\mathrm{l}}$, we used the ANNz2 \phzs\ of the KiDS-Bright foreground galaxy sample. We implemented the contribution of $z_{\mathrm{l}}$ by integrating over the individual redshift probability distributions $p(z_{\mathrm{l}})$ of each lens. This method is shown to be accurate in \cite{Brouwer21}. 
The lensing kernel is wide and therefore the results are not sensitive to the small wings in the lens redshift probability distributions (see Sect.~\ref{Sec:photo-z-model}).
We can thus safely assume that $p(z_{\mathrm{l}})$ can be described by a normal distribution centered at the lens’s \phz, with a standard deviation of $\sigma_{\mathrm{z}} / (1+z_{\mathrm{l}}) = 0.018$ (see Sect.~\ref{Sec:photo-zs}). This simplification over the better-fitted Lorentzian model (Eq.~\ref{Eq:gen-Lorentz}) does not induce significant biases in the present analysis as it will only lead to slightly underestimated errorbars of the lensing signal. In the future, as the statistical power of such measurements increases, these types of details might need to be taken into account, however.

For the source redshifts $z_{\mathrm{s}}$, we followed the method used in \citet{Dvornik18} by integrating over the part of the redshift probability distribution $n(z_{\mathrm{s}})$ where $z_{\mathrm{s}} > z_{\mathrm{l}}$. Thus, the ESD can be directly computed in bins of projected distance $r_{\text{p}}$ to the lenses as follows:
\begin{equation}
\label{eq:ESDmeasured}
\Delta \Sigma_{\text{gm}} (r_{\text{p}}) = \left[ \frac{\sum_{\mathrm{ls}}\widetilde{w}_{\mathrm{ls}}\epsilon_{\mathrm{t, s}}\Sigma_{\mathrm{cr, ls}}^{\prime}}{\sum_{\mathrm{ls}}\widetilde{w}_{\mathrm{ls}}} \right] \frac{1}{1+\overline{m}} \, ,\end{equation}
where $\Sigma_{\mathrm{cr, ls}}^{\prime} \equiv 1/ \widetilde\Sigma_{\mathrm{cr, ls}}^{-1}$, the sum is over all source-lens pairs in the distance bin, and
\begin{equation}
\overline{m} = \frac{\sum_{i}w_{i}^{\prime}m_{i}}{\sum_{i}w_{i}^{\prime}} \, \end{equation}
is an average correction to the ESD profile that has to be applied to account for the multiplicative bias $m$ in the \emph{lens}fit shear estimates. 
The sum goes over thin redshift slices for which $m$ was obtained using the method presented in \citet{Kannawadi19}, weighted 
by $w^{\prime} = w_{\rm s}\,D(z_{\rm l},z_{\rm s}) / D(z_{\rm s})$ for a given lens-source sample. The value of $\overline{m}$ is 
around $- 0.014$, independent of the scale at which it was computed.

We note that  the measurements presented here are not corrected for the contamination of the source sample by galaxies that are physically associated with the lenses (the so-called boost correction). The impact on $\Delta\Sigma$ is minimal, however, as a result of the weighting with the inverse square of the critical surface density in Eqn.~\eqref{eq:crit_effective} \citep[see for instance the bottom panel of fig.~A4 in][]{Dvornik17}. We also did not subtract the signal around random points, which suppresses large-scale systematics and sample variance \citep{Singh17,Dvornik18}. This improves the robustness of the measurements on scales above $2 h^{-1}\, \mathrm{Mpc}$ \citep{Dvornik18}, which are not particularly relevant in constraining the halo model and halo occupation distribution parameters, and mostly affect the bias present in the 2-halo term, which we do not consider here (see Sect.~\ref{Sec:SHMR}).

\subsection{Comparison with lenses from GAMA}
\label{sec:GGLGAMA}

\begin{figure}
        \includegraphics[width=\columnwidth]{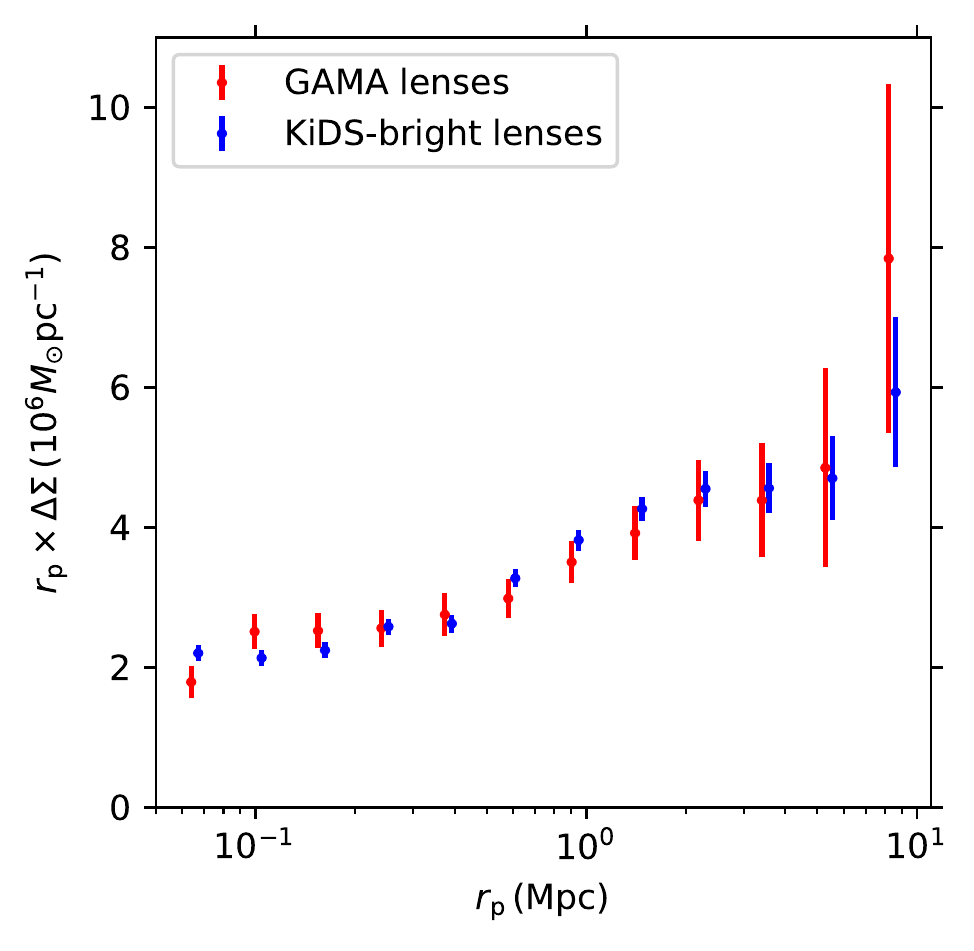}
        \caption{Stacked excess surface density profiles, $\Delta\Sigma$ (multiplied by the distance from the lens $r_\mrm{p}$ in megaparsecs), around lenses with $\log(M_{\star}/M_{\sun}) \in [10.5,11.0]$. The red points show results for $68\,000$ lenses selected from GAMA, whereas the blue points show the signal around $352\,000$ lenses from the KiDS-Bright sample. The KiDS measurements were shifted slightly to the right for clarity.}
        \label{fig:DeltaSigmaComp}
\end{figure}

As a first demonstration of the statistical power of the KiDS-Bright sample for GGL measurements, and to verify the quality of our photometrically selected lens sample, we
directly compared the stacked excess surface density profile,
$\Delta\Sigma$, with that of lenses extracted from GAMA.
For the comparison, we used the stellar masses from the two respective surveys and defined a bin of 0.5 dex around the mode of the $\log M_\star$ distribution, which in both cases is $\sim 10.75$. This selection of $10.5\leq\log(M_{\star}/M_{\sun})\leq11.0$ gives about 68\,000 galaxies in GAMA and 352\,000 in KiDS-Bright; in both cases, this is $\sim35\%$ of the full sample. The resulting excess surface density $\Delta\Sigma$, multiplied by the projected distance from the lens $r_\mrm{p}$ to enhance the large-scale signal, is presented in Fig.~\ref{fig:DeltaSigmaComp} as a function of $r_\mrm{p}$.

The two measurements agree remarkably well, demonstrating that our \phzs\ are sufficient for GGL studies. The small differences in the central values in Fig.~\ref{fig:DeltaSigmaComp} most likely arise from the inclusion of the whole KiDS-South area to the lensing study.
The reduction in uncertainties also agrees with our expectation: for all scales, $\delta\, \Delta\Sigma_{\mathrm{GAMA}} / \delta\, \Delta\Sigma_{\mathrm{KiDS}} \approx 2.4$, which reflects the 
fact that the KiDS-Bright sample contains $\sim5.6\times$ more galaxies. We also tested how much statistical power we lose by using \phzs. For this we extracted the lensing signal in the same way as for GAMA, namely using the point estimate of the redshift, without its uncertainty (by dropping the integration over $p(z_{\mathrm{l}})$ in Eqn.~\ref{eq:crit_effective}). We found that the statistical power is worsened by only $\sim5\%$ when propagating the redshift uncertainty through to the final lensing signal stack. 

The precision improves slightly when the data for the full survey area (1350 deg$^2$) are included. This will make it possible to revisit the earlier study by \cite{Brouwer18} of the lensing signal of "troughs" and "ridges" in the density field of KiDS galaxies, based on the much smaller catalog derived by \citetalias{Bilicki18}. The sample we present in this paper has already been used in other analyses. \cite{Brouwer21} selected isolated galaxies to measure the radial gravitational acceleration around them based on weak lensing measurements, thus extending the so-called radial acceleration relation into the low acceleration regime beyond the outskirts of the observable galaxies. The sample was also used by \cite{Johnston20} as a test-bed for a new method to mitigate observational systematics in angular clustering measurements, in which self-organizing maps are taught the multivariate relationships between observed galaxy number density and systematic tracer variables. This was then used to create corrective random catalogs with spatially variable number densities, mimicking the systematic density modes in the data.

\begin{figure*}
        \includegraphics[width=\textwidth]{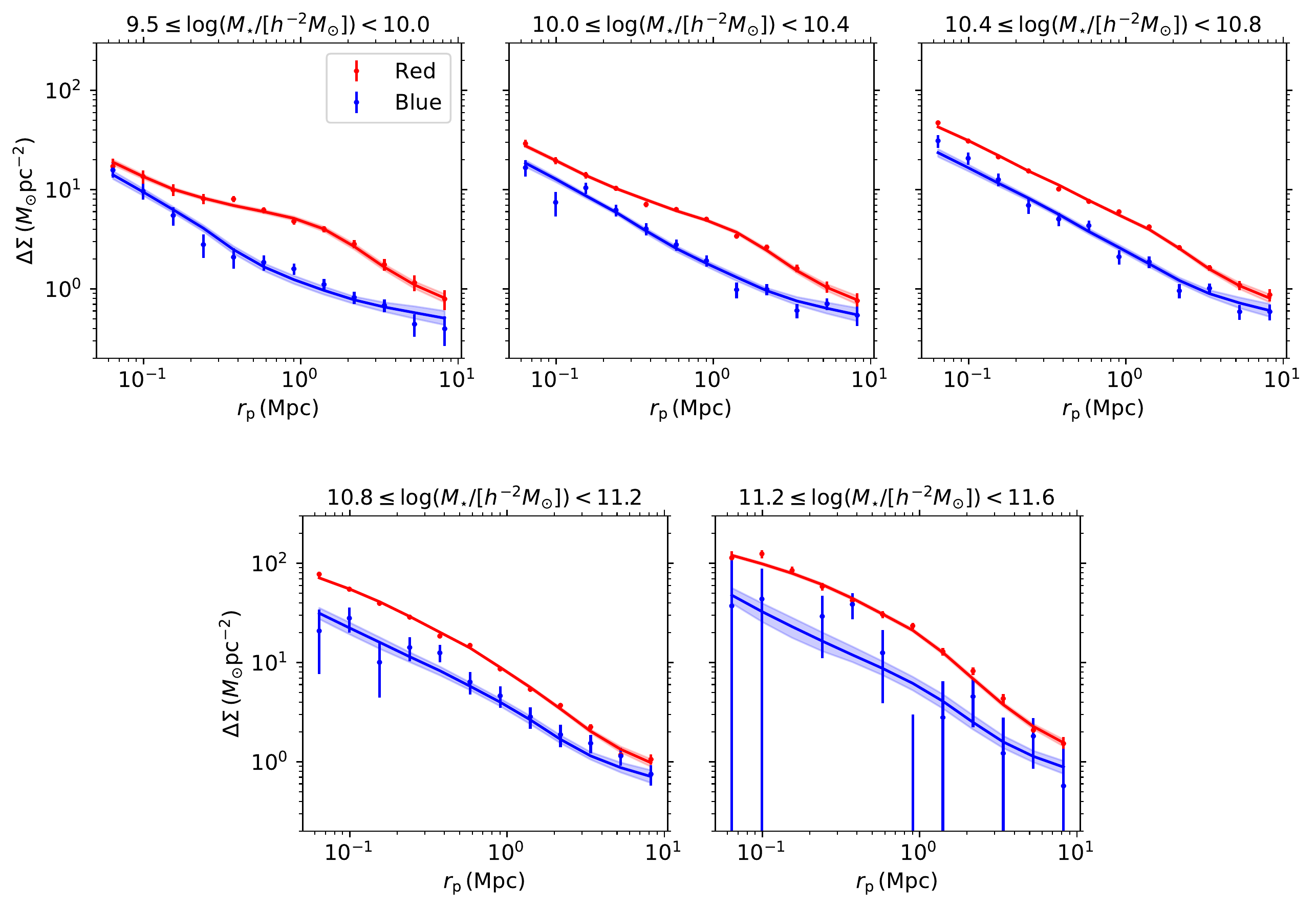}
        \caption{Stacked excess surface density profiles, $\Delta\Sigma$, of the red and blue lenses (points in corresponding colors) in our KiDS-Bright sample in the four stellar mass bins labeled at the top. The lines
        indicate the best-fitting halo model, with contributions from both centrals and satellites (red and blue lines with shaded bands enclosing the 68\% credible intervals). We note that the model was fit to all stellar mass bins simultaneously, but separately for the red and blue populations.}
        \label{fig:DeltaSigma}
\end{figure*}

The improvement in statistical power will also allow for better constraints on the halo model and the associated halo occupation properties. The small-scale measurements accessible with such a sample will provide better constraints on the galaxy bias in the nonlinear regime and allow us to test our assumption about the validity of the halo model. Finally, we anticipate that this kind of wide-angle lens sample can improve cosmological constraints from multi-probe analyses employing GGL.

\subsection{Stellar-to-halo-mass relation}
\label{Sec:SHMR}

As a further demonstration of the quality of our data, we used the KiDS-Bright sample to explore the stellar-to-halo-mass relation (SHMR) for the blue and red galaxies separately. Earlier GGL studies have shown that these differ \citep[e.g.,][]{Hoekstra05,Velander14, Mandelbaum16}, which is also seen in hydrodynamic simulations \citep[e.g.,][]{Correa20}. Nonetheless there is no consensus in the literature yet because other approaches have arrived at different conclusions \citep[see][for a detailed overview and discussion]{Wechsler18}. Some of the differences may arise from the stellar mass estimates and the specific selection of the subsamples. For this reason we do not compare our findings to the literature, but defer such a detailed comparison to future work. Our aim is merely to demonstrate the potential of our data for studies of the SHMR.

We split the KiDS-Bright sample by color using the cut defined in Sect.~\ref{Sec:stellar_masses} (see Fig.~\ref{fig:red-blue-split}). We selected lenses with $\zph>0.1$ and used our stellar mass estimates to subdivide the blue and red galaxies into five stellar mass intervals, with the bin edges: $\smash{\log\left(M_{\star}/[h^{-2}M_{\sun}]\right)} = \{9.5,10.0,10.4,10.8,11.2,11.6\}$. In this section we give results in terms of an explicitly $h$-dependent mass unit, as used in our modeling, rather than adopting the specific value $h=0.7$, as used elsewhere. The properties of the subsamples are reported in Table \ref{tab:stellar-mass-bins}. For each stellar mass bin of the two color selections, we measured the lensing signal as described above, and the results are shown in Fig.~\ref{fig:DeltaSigma}. For all subsamples we detect a significant signal, demonstrating the value of our bright galaxy selection.

\begin{table}
        \caption{Overview of the number of lens galaxies, median stellar masses of the galaxies, and median redshifts in each selected mass bin.} 
        \label{tab:stellar-mass-bins}
        \centering
        \begin{tabular}{lrrrr} 
                \toprule
                Bin & \multicolumn{1}{c}{$\log M_\star$ range} & \multicolumn{1}{c}{$N_{\mathrm{red}}$} & \multicolumn{1}{c}{$\log M_{\star, \mathrm{med}}^\mrm{(red)}$} & \multicolumn{1}{c}{$z_{\mathrm{med}}^\mrm{(red)}$} \\ 
                \midrule
                1 & [9.5,10.0)  & 52\,813  & 9.83 & 0.16 \\
                2 & [10.0,10.4) & 119\,038  & 10.23 & 0.23 \\
                3 & [10.4,10.8) & 147\,342  & 10.58 & 0.29 \\
                4 & [10.8,11.2) & 52\,320  & 10.92 & 0.36 \\
                5 & [11.2,11.6) & 4\,342   & 11.28 & 0.43 \\
                \midrule
                Bin & \multicolumn{1}{c}{$\log M_\star$ range} & \multicolumn{1}{c}{$N_{\mathrm{blue}}$} & \multicolumn{1}{c}{$\log M_{\star, \mathrm{med}}^\mrm{(blue)}$} & \multicolumn{1}{c}{$z_{\mathrm{med}}^\mrm{(blue)}$} \\ 
                \midrule
                1 & [9.5,10.0) & 97\,786 & 9.75 & 0.22 \\
                2 & [10.0,10.4) & 85\,594 & 10.20 & 0.29 \\
                3 & [10.4,10.8) & 60\,541 & 10.55 & 0.36 \\
                4 & [10.8,11.2) & 8\,839  & 10.88 & 0.40 \\
                5 & [11.2,11.6) & 428     & 11.31 & 0.41 \\
                
                \bottomrule
        \end{tabular}
        \tablefoot{Stellar masses are given in units of $\smash{\log\left(M_{\star} / [h^{-2}M_{\sun}]\right)}$. The median stellar masses are used as an estimate of the stellar contribution to the total lensing signal described as a point-like source.}
\end{table}

To infer the corresponding halo masses, we need to fit a model to the lensing signal. Numerical simulations show that the dark matter distribution in halos is well described by an Navarro-Frenk-White (NFW) profile \citep{NFW97}, but the signals shown in Fig.~\ref{fig:DeltaSigma}, especially those of the red galaxies with low stellar masses, show a more complex dependence with radius. At large radii the lensing signal is enhanced by the clustering of galaxies, whereas on small scales satellite galaxies contribute, causing a wide "bump" around 1~Mpc.

The influence of neighboring galaxies can be reduced by
selecting "isolated" lenses, so that a simple model can still describe the measurements. This approach was used by \cite{Hoekstra05} and \cite{Brouwer21}, but at the expense of significantly reducing the lens sample size. 
Here, inspired by the halo model \citep{Seljak00,CS02}, we estimate the mean halo mass of central galaxies as a function of stellar mass by modeling the contributions of both central and satellite galaxies jointly. The SHMR of central galaxies is parameterized using the following equation:
\begin{equation}\label{eq:CMF4}
M_{\star}(M_{\mathrm{h}}) = M_{0} \frac{(M_{\mathrm{h}}/M_{1})^{\gamma_{1}}}{[1 + (M_{\mathrm{h}}/M_{1})]^{\gamma_{1} - \gamma_{2}}}\,.
\end{equation}
This relation has an intrinsic scatter, and we assume that the distribution of $\log(M_{\star})$ at a fixed halo mass is a Gaussian with a dispersion $\sigma_c$. It is important to include this intrinsic scatter as it enables the model to account for Eddington bias \citep{Leauthaud12, Cacciato13}.

The model itself is based on the halo model implementation presented in \citet{vanUitert16}, but in our case we adopted a separate normalization of the concentration of the dark matter density profile for central and satellite galaxies, a free normalization of the two-halo term, and a fixed subhalo mass for satellite galaxies. The free parameters that describe the lensing signal around a galaxy with a given mass are thus the following: the normalization of the concentration-mass relation for central galaxies, $f_c$; the normalization of the SHMR,  $M_0$; its characteristic mass scale, $M_1$; the low and high mass end slopes, $\gamma_1$ and $\gamma_2$; and the normalization of the concentration-mass relation for satellite galaxies, $f_s$. We simply fit for the normalization of the 2-halo term, $b$, but did not aim to interpret its value.

The number density of halos of a given stellar mass is not uniform, and this needs to be accounted for in the model. Moreover, in doing so, we need to distinguish between central and satellite galaxies because the satellite fraction itself depends on mass. To do so, we used the conditional stellar mass function (CSMF), which we describe in more detail in Appendix~\ref{App:hod}. This introduces the following additional parameters: the high mass slope of the Schechter function, $\alpha_s$, and the free parameters for the normalization of the Schechter function used for satellite galaxies, $b_1$ and $b_2$.
Finally, we note that we assume that none of the parameters depend on redshift and that the parameters of the Schechter function are constrained by the lensing signal alone.

The model, as detailed in Appendix~\ref{App:hod}, implicitly assumes that we employed a complete volume-limited sample of lenses. This is not the case here because the cut in apparent $r$-band magnitude leads to incompleteness that is larger for low stellar masses, with the selection of red galaxies affected the most. A proper analysis, which is beyond the scope of our exploratory study, would have to explicitly include the apparent magnitude cut of the sample in the model. This is also required if one would like to jointly model the GGL signal, the stellar mass function, and the clustering signal. 

\begin{table}
    \centering
    \caption{Parameter space ranges and marginalized posterior estimates of the free parameters used in our model, for both the red and blue sample.}
    \label{tab:halo-model-fit}
    \begin{tabular}{lrrr}
        \toprule
        Parameter & Priors & Red & Blue \\
        \midrule
        $f_c$ & $[0, 1]$ & $0.993^{+0.002}_{-0.021}$ & -- \\ \addlinespace
        $\log(M_{0}/[h^{-2}M_{\odot}])$ & $[7,13]$ & $10.39^{+0.14}_{-0.15}$ & $10.11^{+0.980}_{-0.087}$ \\ \addlinespace             $\log(M_{1}/[h^{-2}M_{\odot}])$ & $[9,14]$ & $11.74^{+0.18}_{-0.20}$ & $11.78^{+0.59}_{-0.38}$ \\ \addlinespace
        $\gamma_1$ & ${\cal N}(3,3)$ & $5.0^{+2.2}_{-1.6}$ & $2.1^{+1.0}_{-1.0}$ \\ \addlinespace
        $\gamma_2$ & $[0,10]$ & $0.47^{+0.17}_{-0.14}$ & $0.65^{+0.56}_{-0.50}$ \\ \addlinespace
        $\sigma_c$ & $[0.05,2.5]$ & $0.064^{+0.046}_{-0.014}$ & $0.28^{+0.36}_{-0.18}$ \\ \addlinespace
        $b$ & $[0.2,5]$ & $0.90^{+0.15}_{-0.11}$ & $0.73^{+0.26}_{-0.25}$ \\ \addlinespace
        $f_s$ & $[0, 1]$ & $0.56^{+0.29}_{-0.15}$ & -- \\ \addlinespace
        $\alpha_s$ & ${\cal N}(-1.1,0.9)$ & $-1.286^{+0.121}_{-0.079}$ & $-0.75^{+0.23}_{-0.14}$ \\ \addlinespace
        $b_1$ & ${\cal N}(0.0,1.5)$ & $-0.65^{+0.12}_{-0.14}$ & $-0.42^{+0.41}_{-0.22}$ \\ \addlinespace
        $b_2$ & ${\cal N}(1.5,1.5)$ & $0.97^{+0.18}_{-0.13}$ & $0.63^{+0.81}_{-0.61}$ \\
        \bottomrule
    \end{tabular}
    \tablefoot{$M_0$ is the normalization of the stellar-to-halo mass relation (SHMR); $M_1$ is the characteristic mass scale of the same SHMR; $f_c$ is the normalization of the concentration-mass relation; $\sigma_c$ is the scatter between the stellar and halo mass; $\gamma_1$ and $\gamma_2$ are the low and high-mass slopes of the SHMR; $f_s$ is the normalization of the concentration–mass relation for satellite galaxies; and $\alpha_s$, $b_0$, and $b_1$ govern the behavior of the satellite galaxies. As the parameters $f_c$ and $f_s$ of the blue sample are not constrained and they recover the prior ranges, we do not provide their values. As discussed in the text, the parameters that describe the CSMF are biased as a result of the cut in apparent magnitude that defines the KiDS-Bright sample.}
\end{table}

The observed lensing signal is, however, most sensitive to the average halo mass of the sample of lenses, so that the resulting mean SHMR for central galaxies is expected to be close to the true one. We stress, however, that the parameters that describe the CSMF will be biased. To test this expectation, we
examined how the magnitude cut changes the stellar mass and halo mass distributions. We used the MICEv2 simulations\footnote{\url{http://maia.ice.cat/mice/}} \citep{Carretero15,Crocce15, Fosalba15a,Fosalba15b} to select central galaxies with $0.1<z<0.5$, which we split into blue and red samples. We used the definitions of stellar mass bins listed in Table~\ref{tab:stellar-mass-bins} and computed the corresponding mean stellar and halo masses. We also repeated the measurements, after we applied a cut in apparent magnitude, $m_r<20$, to mimic the selection of the KiDS-Bright sample. As expected, the resulting stellar mass functions are biased low for low stellar masses, with the red galaxies being affected the most. In contrast, the changes in the mean SHMR are small: the mean $\log(M_\star)$ is less than 0.05 dex lower and the intrinsic scatter is not significantly affected either. Given the uncertainties in the stellar masses themselves, we therefore conclude that the magnitude cut has a negligible impact on the inferred SHMR. Nonetheless, we defer a quantitative interpretation of the results to future work.

We fit our model (see Appendix~\ref{App:hod} for a summary) to the lensing signal of each of the color-selected subsamples (that is, a single model for all the stellar mass bins). 
The priors that we used are listed in Table~\ref{tab:halo-model-fit}. Most priors are flat in the given ranges; the instances with a Gaussian prior are indicated as ${\cal N}(\bar{x}, \sigma(x))$, with mean $\bar{x}$ and a standard deviation $\sigma(x)$.  In the fit we used the bootstrap covariance matrix measured directly on the data \citep[for details see][]{Viola15, Dvornik18}, with the 
correction from \citet{Hartlap07} applied to account for noise in the covariance matrix. 

\begin{figure}
    \centering
        \includegraphics[width=0.9\columnwidth]{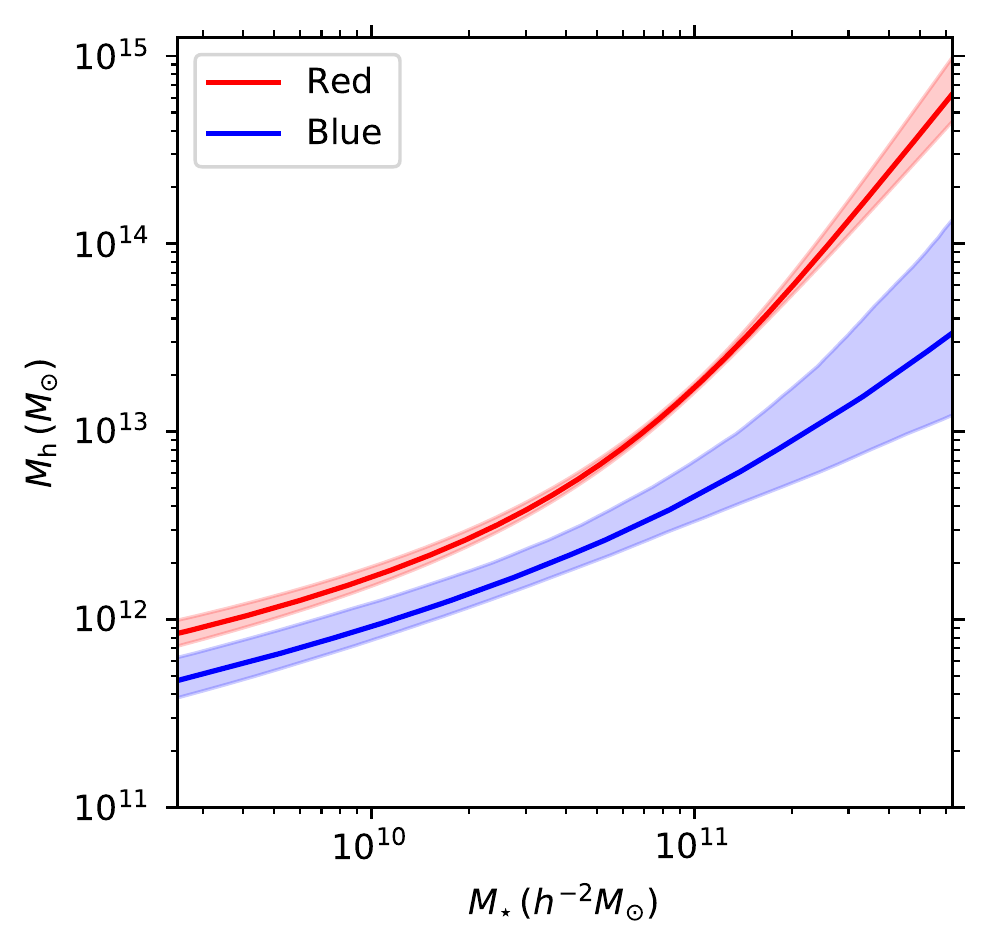}
        \caption{Predicted halo mass as a function of the stellar mass for red and blue galaxies from this study, using the halo model parameters listed in Table~\ref{tab:halo-model-fit}.}
        \label{fig:smhm}
\end{figure}

The best-fit parameters obtained with the Markov chain Monte Carlo method \citep{emcee} for the halo model are reported in Table~\ref{tab:halo-model-fit}, and we show the corresponding models in Fig.~\ref{fig:DeltaSigma} as lines, with shaded areas indicating the uncertainty. The reduced $\chi^2_{\mathrm{red}}$ of the halo model fit is $1.92$ and $1.91$ for the red and blue samples, respectively, with 48 degrees of freedom. Although the $\chi^2_{\mathrm{red}}$ values are high, we note that our model is only an effective description of the signal; our small statistical uncertainties may already point to the need to improve the modeling itself \citep[e.g.,][]{MV20,Sugiyama20}. The best-fit SHMR models and their uncertainties for the red and blue samples are shown in Fig.~\ref{fig:smhm}. The data cannot constrain the concentration normalization of blue central and satellite galaxies, and for these parameters we recover their prior ranges.

Our lensing results suggest that red galaxies with observed stellar masses of $M_\star<5\times 10^{10}h^{-2}M_\odot$ occupy dark matter halos that are about a factor of two more massive than those of blue galaxies with similar stellar masses. At the high mass end, however, the differences are larger and red galaxies at a given stellar mass are found in much more massive halos. Qualitatively, these results are in good agreement with the bimodality found by \citet{Mandelbaum16}. 

The accuracy of the stellar mass estimates from SED modeling suffer from systematic uncertainties, arising from assumptions about the star formation history, the initial mass function, or the photometry itself. Although our split by rest-frame color might exacerbate such systematics, the difference we observe is too large to be solely attributed to them. Nonetheless,  a more detailed investigation is needed before we can quantify the various sources of bias more reliably. Moreover, as discussed above, our model does not fully capture the impact of the magnitude limit of the KiDS-Bright sample. Similarly, a quantitative comparison with previous results \citep[e.g.,][]{Velander14, Mandelbaum16} requires a careful replication of their sample selections and stellar mass determination.

\section{Conclusions and future prospects}
\label{Sec:conclusions_and_prospects}

We selected a sample of bright galaxies using the 9-band photometry from KiDS DR4 \citep{KiDS-DR4} that closely resembles the highly complete spectroscopic dataset from GAMA \citep{GAMA}. For an optimal completeness-purity trade-off, we applied a KiDS magnitude limit of $r_\mrm{auto} < 20$ and employed three star-galaxy separation criteria based on KiDS photometry. This resulted in a highly pure sample of galaxies that matches the properties of GAMA very well, with only $\sim1\%$ of the KiDS-Bright galaxies not represented
with respect to GAMA. The dataset probes the large-scale structure at a mean redshift of $\meanz \simeq 0.23$ and reaches up to $z \lesssim 0.5$, although with decreasing completeness at these high redshifts due to its flux-limited character.

The very good match between the two samples allowed us to take full advantage of supervised machine learning regression and derive statistically accurate and precise photometric redshifts for the entire KiDS-Bright catalog. To do so, we used artificial neural networks implemented in the ANNz2 package \citep{ANNz2}. The resulting \phzs\ have a small scatter of $\sigma_z\sim0.018(1+z)$ and a mean bias $|\meandz| <10^{-3}$. The \phz\ performance does not depend on the $r$-band magnitude nor on the \phz\ for $0.05 < \zph < 0.5$.
The \phz\ error distribution is close to Gaussian, but a generalized Lorentzian captures the 
slightly broader wings better. 

We exploited the nine-band coverage and the high-quality \phzs\
to derive robust absolute magnitudes, rest-frame colors, and stellar masses using the \textsc{LePhare} SED-fitting tool \citep{Arnouts99}. We employed these derivations to split the sample into red and blue galaxies based on the rest-frame $u-g$ color and absolute $r$-band magnitude. The red galaxies have better \phzs\ than the full sample, with $\sigma_z \sim 0.015(1+z)$ at the mean redshift $\meanz \sim 0.27$. Nonetheless, the \phzs\ for the blue galaxies
are also excellent with $|\meandz| \sim 10^{-3}$ and $\sigma_z \sim 0.019(1+z)$). This exquisite performance is achieved thanks to the very complete coverage of the GAMA training set, which is free of any color preselections.

A comparison of the stellar masses with independent estimates from GAMA \citep{Taylor11,GAMA-LAMBDAR} shows excellent agreement, with $\Delta \log M_\star \equiv \log M^\mrm{KiDS}_* - \log M^\mrm{GAMA}_* = -0.09 \pm0.18$ dex (median and SMAD). Our use of photometric redshifts accounts for 9\% of this scatter, demonstrating the sample's potential for scientific exploitation.
As a scientific verification of the KiDS-Bright dataset, we measured the galaxy-galaxy lensing signal for galaxies
with stellar masses in the range $10.5 \leq \log M_\star/M_\sun \leq 11$ 
and compared these directly to a similar selection using GAMA only. The lensing signals agree over two decades in angular separation, while the uncertainties are a factor of $\sim2.4$ smaller for the sample of KiDS-Bright lenses.

Motivated by this agreement, we measured the lensing signal around the blue and red galaxies in five stellar mass bins, ranging from  $\smash{\log\left( M_\star/h^{-2}M_\sun\right)}=9.5$ to $11.6$, and we detected  significant signals in all cases. 
The measurements were fitted with a model that includes both central and satellite galaxies \citep[e.g.,][]{Dvornik18}. Their relative contributions as a function of stellar mass are described using a conditional stellar mass function. The resulting parameters, however, are biased, because the KiDS-Bright magnitude limit leads to incompleteness at low stellar masses, with the red sample being affected the most. Fortunately, a comparison to a simulated catalog of galaxies from MICEv2 suggests that the SHMR is not affected significantly.

We used this model to constrain the SHMR for blue and red galaxies separately. We find that blue and red galaxies with observed stellar masses of $M_\star<5\times 10^{10}h^{-2}M_\odot$ occupy dark matter halos that are about a factor two more massive than those of blue galaxies with similar stellar masses. For
stellar masses of $M_\star \gtrsim 10^{11}h^{-2}M_\odot$, the model predicts however that the dark matter halos of red galaxies  are much more massive than those of blue galaxies with the same stellar mass. This result is in good qualitative agreement with similar findings by \cite{Mandelbaum16}. A more detailed comparison, however, is beyond the scope of this paper because it would also require a careful comparison of the stellar masses, whilst accounting for differences in the sample selection.

Our results demonstrate the value of combining highly complete spectroscopy with high-quality imaging data. In the coming decade, further advances will be made on both fronts. Large spectroscopic surveys will probe both larger volumes and fainter galaxies than current wide-angle redshift catalogs, to the benefit of the already existing imaging surveys. In the case of KiDS, further improvements will be possible thanks to new overlapping complete redshift samples deeper than GAMA, such as the ongoing Deep Extragalactic VIsible Legacy Survey \citep[DEVILS,][]{DEVILS} that aims at a very complete selection with a flux limit of $Y < 21.2$ in fields that partly overlap with KiDS imaging. On a longer timescale, the 4-metre Multi-Object Spectroscopic Telescope \citep[4MOST,][]{4MOST} should deliver denser redshift coverage than GAMA over the full KiDS area, in particular from its  Wide-Area VISTA Extragalactic \citep[WAVES,][]{WAVES} and Cosmology Redshift Surveys \citep{4MOST-CRS}.
 
Such larger and deeper spectroscopic data will be ideally suited to exploit Stage-IV imaging surveys, such as the Rubin Observatory's Legacy Survey of Space and Time \citep{LSST} and {\it Euclid} \citep{Euclid}. Those will cover areas more than $10\times$ larger at a greater depth than the Stage-III surveys such as KiDS. The resulting increase in the statistical power will, however, require much better handling of systematics, starting from those in the selection of lenses for GGL and 3$\times$2pt analyses. Our study demonstrates that one possible approach toward this goal is to extract a well-characterized, flux-limited galaxy catalog, provided that a matched spectroscopic subsample is available to calibrate this selection and to estimate robust photometric redshifts. Such samples can be enhanced with deeper, yet less complete, photometric selections of luminous red galaxies \citep[e.g.,][]{redMaGiC,Vakili20} and adaptive magnitude cuts as a function of \phz\ \citep{Porredon20} to probe a larger range of lens redshifts and luminosities.

\begin{acknowledgements}
We thank the referee Gabriel Brammer for valuable suggestions on the paper and John Peacock for useful comments on the earlier version of the manuscript.\\

Based on data products from observations made with ESO Telescopes at the La Silla Paranal Observatory under program IDs 177.A-3016, 177.A-3017 and 177.A-3018, and on data products produced by Target/OmegaCEN, INAF-OACN, INAF-OAPD and the KiDS production team, on behalf of the KiDS consortium. OmegaCEN and the KiDS production team acknowledge support by NOVA and NWO-M grants. Members of INAF-OAPD and INAF-OACN also acknowledge the support from the Department of Physics \& Astronomy of the University of Padova, and of the Department of Physics of Univ. Federico II (Naples).\\

GAMA is a joint European-Australasian project based around a spectroscopic campaign using the Anglo-Australian Telescope. The GAMA input catalog is based on data taken from the Sloan Digital Sky Survey and the UKIRT Infrared Deep Sky Survey. Complementary imaging of the GAMA regions is being obtained by a number of independent survey programs including GALEX MIS, VST KiDS, VISTA VIKING, WISE, Herschel-ATLAS, GMRT, and ASKAP providing UV to radio coverage. GAMA is funded by the STFC (UK), the ARC (Australia), the AAO, and the participating institutions. The GAMA website is \url{http://www.gama-survey.org/}.\\

MBi is supported by the Polish National Science Center through grants no. 2020/38/E/ST9/00395, 2018/30/E/ST9/00698 and 2018/31/G/ST9/03388. MBi and SJN acknowledge the support from the Polish Ministry of Science and Higher Education through grant DIR/WK/2018/12.
AD, AHW and HHi are supported by an European Research Council Consolidator Grant (No. 770935).
HHo acknowledges support from the Netherlands Organisation for Scientific Research (NWO) through grant 639.043.512.
This work is part of the Delta ITP consortium, a program of the Netherlands Organisation for Scientific Research (NWO) that is funded by the Dutch Ministry of Education, Culture and Science (OCW).
MA, BG and CH acknowledge support from the European Research Council under grant number 647112.
BG is supported by the Royal Society through an Enhancement Award (RGF/EA/181006).
CH acknowledges support from the Max Planck Society and the Alexander von Humboldt Foundation in the framework of the Max Planck-Humboldt Research Award endowed by the Federal Ministry of Education and Research. 
HHi is supported by a Heisenberg grant of the Deutsche Forschungsgemeinschaft (Hi 1495/5-1).
KK acknowledges support from the Royal Society and Imperial College,
SJN is supported by the Polish National Science Center through grant UMO-2018/31/N/ST9/03975.
HYS acknowledges the support from NSFC of China under grant 11973070, the Shanghai Committee of Science and Technology grant No.19ZR1466600 and Key Research Program of Frontier Sciences, CAS, Grant No. ZDBS-LY-7013.
\\ 

This work has made use of \textsc{TOPCAT} \citep{TOPCAT} and \textsc{STILTS} \citep{STILTS} software, as well as of \textsc{python} (\url{www.python.org}), including the packages \textsc{NumPy} \citep{NumPy}, \textsc{SciPy} \citep{SciPy}, and \textsc{Matplotlib} \citep{Matplotlib}.\\

\textit{Author contributions:} All authors contributed to the development and writing of this paper. The authorship list is given in two groups: the lead authors (MB, AD, HHo, AHW, NEC, MV), followed by an alphabetical group which includes  those who have either made a significant contribution to the data products, or to the scientific analysis.

\end{acknowledgements}



\bibliographystyle{mnras}
\bibliography{KiDS-DR4-bright}

\begin{thebibliography}{}
\makeatletter
\relax
\def\mn@urlcharsother{\let\do\@makeother \do\$\do\&\do\#\do\^\do\_\do\%\do\~}
\def\mn@doi{\begingroup\mn@urlcharsother \@ifnextchar [ {\mn@doi@}
  {\mn@doi@[]}}
\def\mn@doi@[#1]#2{\def\@tempa{#1}\ifx\@tempa\@empty \href
  {http://dx.doi.org/#2} {doi:#2}\else \href {http://dx.doi.org/#2} {#1}\fi
  \endgroup}
\def\mn@eprint#1#2{\mn@eprint@#1:#2::\@nil}
\def\mn@eprint@arXiv#1{\href {http://arxiv.org/abs/#1} {{\tt arXiv:#1}}}
\def\mn@eprint@dblp#1{\href {http://dblp.uni-trier.de/rec/bibtex/#1.xml}
  {dblp:#1}}
\def\mn@eprint@#1:#2:#3:#4\@nil{\def\@tempa {#1}\def\@tempb {#2}\def\@tempc
  {#3}\ifx \@tempc \@empty \let \@tempc \@tempb \let \@tempb \@tempa \fi \ifx
  \@tempb \@empty \def\@tempb {arXiv}\fi \@ifundefined
  {mn@eprint@\@tempb}{\@tempb:\@tempc}{\expandafter \expandafter \csname
  mn@eprint@\@tempb\endcsname \expandafter{\@tempc}}}

\bibitem[\protect\citeauthoryear{{Abazajian} et~al.,}{{Abazajian}
  et~al.}{2009}]{SDSS.DR7}
{Abazajian} K.~N.,  et~al., 2009, \mn@doi [\apjs]
  {10.1088/0067-0049/182/2/543}, \href
  {https://ui.adsabs.harvard.edu/abs/2009ApJS..182..543A} {182, 543}

\bibitem[\protect\citeauthoryear{{Abbott} et~al.,}{{Abbott}
  et~al.}{2018}]{Abbott18}
{Abbott} T.~M.~C.,  et~al., 2018, \mn@doi [\prd] {10.1103/PhysRevD.98.043526},
  \href {https://ui.adsabs.harvard.edu/abs/2018PhRvD..98d3526A} {98, 043526}

\bibitem[\protect\citeauthoryear{{Abolfathi} et~al.,}{{Abolfathi}
  et~al.}{2018}]{SDSS.DR14}
{Abolfathi} B.,  et~al., 2018, \mn@doi [\apjs] {10.3847/1538-4365/aa9e8a},
  \href {https://ui.adsabs.harvard.edu/abs/2018ApJS..235...42A} {235, 42}

\bibitem[\protect\citeauthoryear{{Aihara} et~al.,}{{Aihara} et~al.}{2018}]{HSC}
{Aihara} H.,  et~al., 2018, \mn@doi [\pasj] {10.1093/pasj/psx066}, \href
  {http://adsabs.harvard.edu/abs/2018PASJ...70S...4A} {70, S4}

\bibitem[\protect\citeauthoryear{{Alam} et~al.,}{{Alam} et~al.}{2017}]{Alam17}
{Alam} S.,  et~al., 2017, \mn@doi [\mnras] {10.1093/mnras/stx721}, \href
  {https://ui.adsabs.harvard.edu/abs/2017MNRAS.470.2617A} {470, 2617}

\bibitem[\protect\citeauthoryear{{Alam} et~al.,}{{Alam} et~al.}{2021}]{eBOSS}
{Alam} S.,  et~al., 2021, \mn@doi [\prd] {10.1103/PhysRevD.103.083533}, \href
  {https://ui.adsabs.harvard.edu/abs/2021PhRvD.103h3533A} {103, 083533}

\bibitem[\protect\citeauthoryear{{Alarcon} et~al.,}{{Alarcon}
  et~al.}{2021}]{Alarcon20}
{Alarcon} A.,  et~al., 2021, \mn@doi [\mnras] {10.1093/mnras/staa3659}, \href
  {https://ui.adsabs.harvard.edu/abs/2021MNRAS.501.6103A} {501, 6103}

\bibitem[\protect\citeauthoryear{{Amon} et~al.,}{{Amon} et~al.}{2018}]{Amon18}
{Amon} A.,  et~al., 2018, \mn@doi [\mnras] {10.1093/mnras/sty1624}, \href
  {https://ui.adsabs.harvard.edu/abs/2018MNRAS.479.3422A} {479, 3422}

\bibitem[\protect\citeauthoryear{{Arnouts}, {Cristiani}, {Moscardini},
  {Matarrese}, {Lucchin}, {Fontana}  \& {Giallongo}}{{Arnouts}
  et~al.}{1999}]{Arnouts99}
{Arnouts} S.,  {Cristiani} S.,  {Moscardini} L.,  {Matarrese} S.,  {Lucchin}
  F.,  {Fontana} A.,   {Giallongo} E.,  1999, \mn@doi [\mnras]
  {10.1046/j.1365-8711.1999.02978.x}, \href
  {https://ui.adsabs.harvard.edu/abs/1999MNRAS.310..540A} {310, 540}

\bibitem[\protect\citeauthoryear{{Balaguera-Antol{\'\i}nez}, {Bilicki},
  {Branchini}  \& {Postiglione}}{{Balaguera-Antol{\'\i}nez}
  et~al.}{2018}]{BA18}
{Balaguera-Antol{\'\i}nez} A.,  {Bilicki} M.,  {Branchini} E.,   {Postiglione}
  A.,  2018, \mn@doi [\mnras] {10.1093/mnras/sty262}, \href
  {https://ui.adsabs.harvard.edu/abs/2018MNRAS.476.1050B} {476, 1050}

\bibitem[\protect\citeauthoryear{{Baldry} et~al.,}{{Baldry}
  et~al.}{2010}]{GAMA-input}
{Baldry} I.~K.,  et~al., 2010, \mn@doi [\mnras]
  {10.1111/j.1365-2966.2010.16282.x}, \href
  {https://ui.adsabs.harvard.edu/abs/2010MNRAS.404...86B} {404, 86}

\bibitem[\protect\citeauthoryear{{Baldry} et~al.,}{{Baldry}
  et~al.}{2012}]{Baldry12}
{Baldry} I.~K.,  et~al., 2012, \mn@doi [\mnras]
  {10.1111/j.1365-2966.2012.20340.x}, \href
  {https://ui.adsabs.harvard.edu/abs/2012MNRAS.421..621B} {421, 621}

\bibitem[\protect\citeauthoryear{{Baldry} et~al.,}{{Baldry}
  et~al.}{2018}]{GAMA-DR3}
{Baldry} I.~K.,  et~al., 2018, \mn@doi [\mnras] {10.1093/mnras/stx3042}, \href
  {https://ui.adsabs.harvard.edu/abs/2018MNRAS.474.3875B} {474, 3875}

\bibitem[\protect\citeauthoryear{{Baldry}, {Sullivan}, {Rani}  \&
  {Turner}}{{Baldry} et~al.}{2021}]{Baldry21}
{Baldry} I.~K.,  {Sullivan} T.,  {Rani} R.,   {Turner} S.,  2021, \mn@doi
  [\mnras] {10.1093/mnras/staa3327}, \href
  {https://ui.adsabs.harvard.edu/abs/2021MNRAS.500.1557B} {500, 1557}

\bibitem[\protect\citeauthoryear{Bartelmann \& Schneider}{Bartelmann \&
  Schneider}{2001}]{Bartelmann1999}
Bartelmann M.,  Schneider P.,  2001, \mn@doi [Phys. Rep.]
  {10.1016/S0370-1573(00)00082-X}, 340, 291

\bibitem[\protect\citeauthoryear{{Beck}, {Szapudi}, {Flewelling}, {Holmberg},
  {Magnier}  \& {Chambers}}{{Beck} et~al.}{2021}]{Beck21}
{Beck} R.,  {Szapudi} I.,  {Flewelling} H.,  {Holmberg} C.,  {Magnier} E.,
  {Chambers} K.~C.,  2021, \mn@doi [\mnras] {10.1093/mnras/staa2587}, \href
  {https://ui.adsabs.harvard.edu/abs/2021MNRAS.500.1633B} {500, 1633}

\bibitem[\protect\citeauthoryear{{Bellstedt} et~al.,}{{Bellstedt}
  et~al.}{2020}]{Bellstedt20}
{Bellstedt} S.,  et~al., 2020, \mn@doi [\mnras] {10.1093/mnras/staa1466}, \href
  {https://ui.adsabs.harvard.edu/abs/2020MNRAS.496.3235B} {496, 3235}

\bibitem[\protect\citeauthoryear{{Ben{\'{\i}}tez}}{{Ben{\'{\i}}tez}}{2000}]{BPZ}
{Ben{\'{\i}}tez} N.,  2000, \mn@doi [\apj] {10.1086/308947}, \href
  {http://adsabs.harvard.edu/abs/2000ApJ...536..571B} {536, 571}

\bibitem[\protect\citeauthoryear{{Bertin} \& {Arnouts}}{{Bertin} \&
  {Arnouts}}{1996}]{SExtractor}
{Bertin} E.,  {Arnouts} S.,  1996, \mn@doi [\aaps] {10.1051/aas:1996164}, \href
  {http://adsabs.harvard.edu/abs/1996A%26AS..117..393B} {117, 393}

\bibitem[\protect\citeauthoryear{{Bilicki}, {Jarrett}, {Peacock}, {Cluver}  \&
  {Steward}}{{Bilicki} et~al.}{2014}]{2MPZ}
{Bilicki} M.,  {Jarrett} T.~H.,  {Peacock} J.~A.,  {Cluver} M.~E.,   {Steward}
  L.,  2014, \mn@doi [\apjs] {10.1088/0067-0049/210/1/9}, \href
  {http://adsabs.harvard.edu/abs/2014ApJS..210....9B} {210, 9}

\bibitem[\protect\citeauthoryear{{Bilicki} et~al.,}{{Bilicki}
  et~al.}{2018}]{Bilicki18}
{Bilicki} M.,  et~al., 2018, \mn@doi [\aap] {10.1051/0004-6361/201731942},
  \href {https://ui.adsabs.harvard.edu/abs/2018A&A...616A..69B} {616, A69}

\bibitem[\protect\citeauthoryear{{Blake} et~al.,}{{Blake}
  et~al.}{2016}]{2dFLenS}
{Blake} C.,  et~al., 2016, \mn@doi [\mnras] {10.1093/mnras/stw1990}, \href
  {http://adsabs.harvard.edu/abs/2016MNRAS.462.4240B} {462, 4240}

\bibitem[\protect\citeauthoryear{{Blake} et~al.,}{{Blake}
  et~al.}{2020}]{Blake20}
{Blake} C.,  et~al., 2020, \mn@doi [\aap] {10.1051/0004-6361/202038505}, \href
  {https://ui.adsabs.harvard.edu/abs/2020A&A...642A.158B} {642, A158}

\bibitem[\protect\citeauthoryear{{Brouwer} et~al.,}{{Brouwer}
  et~al.}{2016}]{Brouwer16}
{Brouwer} M.~M.,  et~al., 2016, \mn@doi [\mnras] {10.1093/mnras/stw1602}, \href
  {http://adsabs.harvard.edu/abs/2016MNRAS.462.4451B} {462, 4451}

\bibitem[\protect\citeauthoryear{{Brouwer} et~al.,}{{Brouwer}
  et~al.}{2017}]{Brouwer17}
{Brouwer} M.~M.,  et~al., 2017, \mn@doi [\mnras] {10.1093/mnras/stw3192}, \href
  {https://ui.adsabs.harvard.edu/abs/2017MNRAS.466.2547B} {466, 2547}

\bibitem[\protect\citeauthoryear{{Brouwer} et~al.,}{{Brouwer}
  et~al.}{2018}]{Brouwer18}
{Brouwer} M.~M.,  et~al., 2018, \mn@doi [\mnras] {10.1093/mnras/sty2589}, \href
  {https://ui.adsabs.harvard.edu/abs/2018MNRAS.481.5189B} {481, 5189}

\bibitem[\protect\citeauthoryear{{Brouwer} et~al.,}{{Brouwer}
  et~al.}{2021}]{Brouwer21}
{Brouwer} M.~M.,  et~al., 2021, \mn@doi [\aap] {10.1051/0004-6361/202040108},
  \href {https://ui.adsabs.harvard.edu/abs/2021A&A...650A.113B} {650, A113}

\bibitem[\protect\citeauthoryear{{Bruzual} \& {Charlot}}{{Bruzual} \&
  {Charlot}}{2003}]{BC03}
{Bruzual} G.,  {Charlot} S.,  2003, \mn@doi [\mnras]
  {10.1046/j.1365-8711.2003.06897.x}, \href
  {https://ui.adsabs.harvard.edu/abs/2003MNRAS.344.1000B} {344, 1000}

\bibitem[\protect\citeauthoryear{{Cacciato}, {van den Bosch}, {More}, {Mo}  \&
  {Yang}}{{Cacciato} et~al.}{2013}]{Cacciato13}
{Cacciato} M.,  {van den Bosch} F.~C.,  {More} S.,  {Mo} H.,   {Yang} X.,
  2013, \mn@doi [\mnras] {10.1093/mnras/sts525}, \href
  {https://ui.adsabs.harvard.edu/abs/2013MNRAS.430..767C} {430, 767}

\bibitem[\protect\citeauthoryear{{Calzetti}, {Kinney}  \&
  {Storchi-Bergmann}}{{Calzetti} et~al.}{1994}]{Calzetti94}
{Calzetti} D.,  {Kinney} A.~L.,   {Storchi-Bergmann} T.,  1994, \mn@doi [\apj]
  {10.1086/174346}, \href
  {https://ui.adsabs.harvard.edu/abs/1994ApJ...429..582C} {429, 582}

\bibitem[\protect\citeauthoryear{{Capaccioli} et~al.,}{{Capaccioli}
  et~al.}{2012}]{Capaccioli12}
{Capaccioli} M.,  et~al., 2012, in Science from the Next Generation Imaging and
  Spectroscopic Surveys. p.~1

\bibitem[\protect\citeauthoryear{{Carretero}, {Castander}, {Gazta{\~n}aga},
  {Crocce}  \& {Fosalba}}{{Carretero} et~al.}{2015}]{Carretero15}
{Carretero} J.,  {Castander} F.~J.,  {Gazta{\~n}aga} E.,  {Crocce} M.,
  {Fosalba} P.,  2015, \mn@doi [\mnras] {10.1093/mnras/stu2402}, \href
  {https://ui.adsabs.harvard.edu/abs/2015MNRAS.447..646C} {447, 646}

\bibitem[\protect\citeauthoryear{{Chabrier}}{{Chabrier}}{2003}]{Chabrier03}
{Chabrier} G.,  2003, \mn@doi [\pasp] {10.1086/376392}, \href
  {https://ui.adsabs.harvard.edu/abs/2003PASP..115..763C} {115, 763}

\bibitem[\protect\citeauthoryear{{Cole} et~al.,}{{Cole} et~al.}{2005}]{Cole05}
{Cole} S.,  et~al., 2005, \mn@doi [\mnras] {10.1111/j.1365-2966.2005.09318.x},
  \href {https://ui.adsabs.harvard.edu/abs/2005MNRAS.362..505C} {362, 505}

\bibitem[\protect\citeauthoryear{{Cooray} \& {Sheth}}{{Cooray} \&
  {Sheth}}{2002}]{CS02}
{Cooray} A.,  {Sheth} R.,  2002, \mn@doi [\physrep]
  {10.1016/S0370-1573(02)00276-4}, \href
  {https://ui.adsabs.harvard.edu/abs/2002PhR...372....1C} {372, 1}

\bibitem[\protect\citeauthoryear{{Correa} \& {Schaye}}{{Correa} \&
  {Schaye}}{2020}]{Correa20}
{Correa} C.~A.,  {Schaye} J.,  2020, \mn@doi [\mnras] {10.1093/mnras/staa3053},
  \href {https://ui.adsabs.harvard.edu/abs/2020MNRAS.499.3578C} {499, 3578}

\bibitem[\protect\citeauthoryear{{Coupon} et~al.,}{{Coupon}
  et~al.}{2015}]{Coupon15}
{Coupon} J.,  et~al., 2015, \mn@doi [\mnras] {10.1093/mnras/stv276}, \href
  {https://ui.adsabs.harvard.edu/abs/2015MNRAS.449.1352C} {449, 1352}

\bibitem[\protect\citeauthoryear{{Crocce}, {Castander}, {Gazta{\~n}aga},
  {Fosalba}  \& {Carretero}}{{Crocce} et~al.}{2015}]{Crocce15}
{Crocce} M.,  {Castander} F.~J.,  {Gazta{\~n}aga} E.,  {Fosalba} P.,
  {Carretero} J.,  2015, \mn@doi [\mnras] {10.1093/mnras/stv1708}, \href
  {https://ui.adsabs.harvard.edu/abs/2015MNRAS.453.1513C} {453, 1513}

\bibitem[\protect\citeauthoryear{{Dalton} et~al.,}{{Dalton}
  et~al.}{2006}]{VIRCAM}
{Dalton} G.~B.,  et~al., 2006, in {McLean} I.~S.,  {Iye} M.,  eds,  Society of
  Photo-Optical Instrumentation Engineers (SPIE) Conference Series Vol. 6269,
  Society of Photo-Optical Instrumentation Engineers (SPIE) Conference Series.
  p. 62690X, \mn@doi{10.1117/12.670018}

\bibitem[\protect\citeauthoryear{{Davies} et~al.,}{{Davies}
  et~al.}{2018}]{DEVILS}
{Davies} L.~J.~M.,  et~al., 2018, \mn@doi [\mnras] {10.1093/mnras/sty1553},
  \href {https://ui.adsabs.harvard.edu/abs/2018MNRAS.480..768D} {480, 768}

\bibitem[\protect\citeauthoryear{{Dawson} et~al.,}{{Dawson}
  et~al.}{2013}]{BOSS}
{Dawson} K.~S.,  et~al., 2013, \mn@doi [\aj] {10.1088/0004-6256/145/1/10},
  \href {http://adsabs.harvard.edu/abs/2013AJ....145...10D} {145, 10}

\bibitem[\protect\citeauthoryear{{Driver} et~al.,}{{Driver}
  et~al.}{2011}]{GAMA}
{Driver} S.~P.,  et~al., 2011, \mn@doi [\mnras]
  {10.1111/j.1365-2966.2010.18188.x}, \href
  {http://adsabs.harvard.edu/abs/2011MNRAS.413..971D} {413, 971}

\bibitem[\protect\citeauthoryear{{Driver} et~al.,}{{Driver}
  et~al.}{2016}]{Driver16}
{Driver} S.~P.,  et~al., 2016, \mn@doi [\mnras] {10.1093/mnras/stv2505}, \href
  {https://ui.adsabs.harvard.edu/abs/2016MNRAS.455.3911D} {455, 3911}

\bibitem[\protect\citeauthoryear{{Driver} et~al.,}{{Driver}
  et~al.}{2019}]{WAVES}
{Driver} S.~P.,  et~al., 2019, \mn@doi [The Messenger]
  {10.18727/0722-6691/5126}, \href
  {https://ui.adsabs.harvard.edu/abs/2019Msngr.175...46D} {175, 46}

\bibitem[\protect\citeauthoryear{{Dvornik} et~al.,}{{Dvornik}
  et~al.}{2017}]{Dvornik17}
{Dvornik} A.,  et~al., 2017, \mn@doi [\mnras] {10.1093/mnras/stx705}, \href
  {http://adsabs.harvard.edu/abs/2017MNRAS.468.3251D} {468, 3251}

\bibitem[\protect\citeauthoryear{{Dvornik} et~al.,}{{Dvornik}
  et~al.}{2018}]{Dvornik18}
{Dvornik} A.,  et~al., 2018, \mn@doi [\mnras] {10.1093/mnras/sty1502}, \href
  {https://ui.adsabs.harvard.edu/abs/2018MNRAS.479.1240D} {479, 1240}

\bibitem[\protect\citeauthoryear{{Eardley} et~al.,}{{Eardley}
  et~al.}{2015}]{Eardley15}
{Eardley} E.,  et~al., 2015, \mn@doi [\mnras] {10.1093/mnras/stv237}, \href
  {https://ui.adsabs.harvard.edu/abs/2015MNRAS.448.3665E} {448, 3665}

\bibitem[\protect\citeauthoryear{{Edge}, {Sutherland}, {Kuijken}, {Driver},
  {McMahon}, {Eales}  \& {Emerson}}{{Edge} et~al.}{2013}]{VIKING}
{Edge} A.,  {Sutherland} W.,  {Kuijken} K.,  {Driver} S.,  {McMahon} R.,
  {Eales} S.,   {Emerson} J.~P.,  2013, The Messenger, \href
  {http://adsabs.harvard.edu/abs/2013Msngr.154...32E} {154, 32}

\bibitem[\protect\citeauthoryear{{Emerson}, {McPherson}  \&
  {Sutherland}}{{Emerson} et~al.}{2006}]{VISTA}
{Emerson} J.,  {McPherson} A.,   {Sutherland} W.,  2006, The Messenger, \href
  {https://ui.adsabs.harvard.edu/abs/2006Msngr.126...41E} {126, 41}

\bibitem[\protect\citeauthoryear{{Erben} et~al.,}{{Erben} et~al.}{2005}]{THELI}
{Erben} T.,  et~al., 2005, \mn@doi [Astronomische Nachrichten]
  {10.1002/asna.200510396}, \href
  {https://ui.adsabs.harvard.edu/abs/2005AN....326..432E} {326, 432}

\bibitem[\protect\citeauthoryear{{Fenech Conti}, {Herbonnet}, {Hoekstra},
  {Merten}, {Miller}  \& {Viola}}{{Fenech Conti} et~al.}{2017}]{FC17}
{Fenech Conti} I.,  {Herbonnet} R.,  {Hoekstra} H.,  {Merten} J.,  {Miller} L.,
    {Viola} M.,  2017, \mn@doi [\mnras] {10.1093/mnras/stx200}, \href
  {https://ui.adsabs.harvard.edu/abs/2017MNRAS.467.1627F} {467, 1627}

\bibitem[\protect\citeauthoryear{{Foreman-Mackey}, {Hogg}, {Lang}  \&
  {Goodman}}{{Foreman-Mackey} et~al.}{2013}]{emcee}
{Foreman-Mackey} D.,  {Hogg} D.~W.,  {Lang} D.,   {Goodman} J.,  2013, \mn@doi
  [\pasp] {10.1086/670067}, \href
  {https://ui.adsabs.harvard.edu/abs/2013PASP..125..306F} {125, 306}

\bibitem[\protect\citeauthoryear{{Fosalba}, {Gazta{\~n}aga}, {Castander}  \&
  {Crocce}}{{Fosalba} et~al.}{2015a}]{Fosalba15b}
{Fosalba} P.,  {Gazta{\~n}aga} E.,  {Castander} F.~J.,   {Crocce} M.,  2015a,
  \mn@doi [\mnras] {10.1093/mnras/stu2464}, \href
  {https://ui.adsabs.harvard.edu/abs/2015MNRAS.447.1319F} {447, 1319}

\bibitem[\protect\citeauthoryear{{Fosalba}, {Crocce}, {Gazta{\~n}aga}  \&
  {Castander}}{{Fosalba} et~al.}{2015b}]{Fosalba15a}
{Fosalba} P.,  {Crocce} M.,  {Gazta{\~n}aga} E.,   {Castander} F.~J.,  2015b,
  \mn@doi [\mnras] {10.1093/mnras/stv138}, \href
  {https://ui.adsabs.harvard.edu/abs/2015MNRAS.448.2987F} {448, 2987}

\bibitem[\protect\citeauthoryear{{Gaia Collaboration} et~al.,}{{Gaia
  Collaboration} et~al.}{2018}]{Gaia-DR2}
{Gaia Collaboration} et~al., 2018, \mn@doi [\aap]
  {10.1051/0004-6361/201833051}, \href
  {https://ui.adsabs.harvard.edu/abs/2018A&A...616A...1G} {616, A1}

\bibitem[\protect\citeauthoryear{{Giblin} et~al.,}{{Giblin}
  et~al.}{2021}]{Giblin20}
{Giblin} B.,  et~al., 2021, \mn@doi [\aap] {10.1051/0004-6361/202038850}, \href
  {https://ui.adsabs.harvard.edu/abs/2021A&A...645A.105G} {645, A105}

\bibitem[\protect\citeauthoryear{{Gunawardhana} et~al.,}{{Gunawardhana}
  et~al.}{2011}]{Gunawardhana11}
{Gunawardhana} M.~L.~P.,  et~al., 2011, \mn@doi [\mnras]
  {10.1111/j.1365-2966.2011.18800.x}, \href
  {https://ui.adsabs.harvard.edu/abs/2011MNRAS.415.1647G} {415, 1647}

\bibitem[\protect\citeauthoryear{{Hang}, {Alam}, {Peacock}  \& {Cai}}{{Hang}
  et~al.}{2021}]{Hang20}
{Hang} Q.,  {Alam} S.,  {Peacock} J.~A.,   {Cai} Y.-C.,  2021, \mn@doi [\mnras]
  {10.1093/mnras/staa3738}, \href
  {https://ui.adsabs.harvard.edu/abs/2021MNRAS.501.1481H} {501, 1481}

\bibitem[\protect\citeauthoryear{{Hartlap}, {Simon}  \& {Schneider}}{{Hartlap}
  et~al.}{2007}]{Hartlap07}
{Hartlap} J.,  {Simon} P.,   {Schneider} P.,  2007, \mn@doi [\aap]
  {10.1051/0004-6361:20066170}, \href
  {https://ui.adsabs.harvard.edu/abs/2007A&A...464..399H} {464, 399}

\bibitem[\protect\citeauthoryear{{Heymans} et~al.,}{{Heymans}
  et~al.}{2021}]{Heymans20}
{Heymans} C.,  et~al., 2021, \mn@doi [\aap] {10.1051/0004-6361/202039063},
  \href {https://ui.adsabs.harvard.edu/abs/2021A&A...646A.140H} {646, A140}

\bibitem[\protect\citeauthoryear{{Hildebrandt} et~al.,}{{Hildebrandt}
  et~al.}{2021}]{Hildebrandt20b}
{Hildebrandt} H.,  et~al., 2021, \mn@doi [\aap] {10.1051/0004-6361/202039018},
  \href {https://ui.adsabs.harvard.edu/abs/2021A&A...647A.124H} {647, A124}

\bibitem[\protect\citeauthoryear{{Hoekstra}, {van Waerbeke}, {Gladders},
  {Mellier}  \& {Yee}}{{Hoekstra} et~al.}{2002}]{Hoekstra02}
{Hoekstra} H.,  {van Waerbeke} L.,  {Gladders} M.~D.,  {Mellier} Y.,   {Yee}
  H.~K.~C.,  2002, \mn@doi [\apj] {10.1086/342228}, \href
  {https://ui.adsabs.harvard.edu/abs/2002ApJ...577..604H} {577, 604}

\bibitem[\protect\citeauthoryear{{Hoekstra}, {Hsieh}, {Yee}, {Lin}  \&
  {Gladders}}{{Hoekstra} et~al.}{2005}]{Hoekstra05}
{Hoekstra} H.,  {Hsieh} B.~C.,  {Yee} H.~K.~C.,  {Lin} H.,   {Gladders} M.~D.,
  2005, \mn@doi [\apj] {10.1086/496913}, \href
  {https://ui.adsabs.harvard.edu/abs/2005ApJ...635...73H} {635, 73}

\bibitem[\protect\citeauthoryear{Hunter}{Hunter}{2007}]{Matplotlib}
Hunter J.~D.,  2007, \mn@doi [Computing in Science Engineering]
  {10.1109/MCSE.2007.55}, 9, 90

\bibitem[\protect\citeauthoryear{{Ilbert} et~al.,}{{Ilbert}
  et~al.}{2006}]{Ilbert06}
{Ilbert} O.,  et~al., 2006, \mn@doi [\aap] {10.1051/0004-6361:20065138}, \href
  {https://ui.adsabs.harvard.edu/abs/2006A&A...457..841I} {457, 841}

\bibitem[\protect\citeauthoryear{{Johnston} et~al.,}{{Johnston}
  et~al.}{2021}]{Johnston20}
{Johnston} H.,  et~al., 2021, \mn@doi [\aap] {10.1051/0004-6361/202040136},
  \href {https://ui.adsabs.harvard.edu/abs/2021A&A...648A..98J} {648, A98}

\bibitem[\protect\citeauthoryear{{Jones} et~al.,}{{Jones}
  et~al.}{2009}]{6dF.DR3}
{Jones} D.~H.,  et~al., 2009, \mn@doi [\mnras]
  {10.1111/j.1365-2966.2009.15338.x}, \href
  {https://ui.adsabs.harvard.edu/abs/2009MNRAS.399..683J} {399, 683}

\bibitem[\protect\citeauthoryear{Jones, Oliphant, Peterson  et~al.}{Jones
  et~al.}{01  }]{SciPy}
Jones E.,  Oliphant T.,  Peterson P.,   et~al., 2001--, {SciPy}: Open source
  scientific tools for {Python}, \url {http://www.scipy.org/}

\bibitem[\protect\citeauthoryear{{Joudaki} et~al.,}{{Joudaki}
  et~al.}{2018}]{Joudaki18}
{Joudaki} S.,  et~al., 2018, \mn@doi [\mnras] {10.1093/mnras/stx2820}, \href
  {http://adsabs.harvard.edu/abs/2018MNRAS.474.4894J} {474, 4894}

\bibitem[\protect\citeauthoryear{{Kannawadi} et~al.,}{{Kannawadi}
  et~al.}{2019}]{Kannawadi19}
{Kannawadi} A.,  et~al., 2019, \mn@doi [\aap] {10.1051/0004-6361/201834819},
  \href {https://ui.adsabs.harvard.edu/abs/2019A&A...624A..92K} {624, A92}

\bibitem[\protect\citeauthoryear{{Kuijken}}{{Kuijken}}{2008}]{GAaP}
{Kuijken} K.,  2008, \mn@doi [\aap] {10.1051/0004-6361:20066601}, \href
  {http://adsabs.harvard.edu/abs/2008A\%26A...482.1053K} {482, 1053}

\bibitem[\protect\citeauthoryear{{Kuijken}}{{Kuijken}}{2011}]{Kuijken11}
{Kuijken} K.,  2011, The Messenger, \href
  {http://adsabs.harvard.edu/abs/2011Msngr.146....8K} {146, 8}

\bibitem[\protect\citeauthoryear{{Kuijken} et~al.,}{{Kuijken}
  et~al.}{2015}]{KiDS-GL}
{Kuijken} K.,  et~al., 2015, \mn@doi [\mnras] {10.1093/mnras/stv2140}, \href
  {http://adsabs.harvard.edu/abs/2015MNRAS.454.3500K} {454, 3500}

\bibitem[\protect\citeauthoryear{{Kuijken} et~al.,}{{Kuijken}
  et~al.}{2019}]{KiDS-DR4}
{Kuijken} K.,  et~al., 2019, \mn@doi [\aap] {10.1051/0004-6361/201834918},
  \href {https://ui.adsabs.harvard.edu/abs/2019A&A...625A...2K} {625, A2}

\bibitem[\protect\citeauthoryear{{LSST Science Collaboration} et~al.,}{{LSST
  Science Collaboration} et~al.}{2009}]{LSST}
{LSST Science Collaboration} et~al., 2009, preprint, \href
  {http://adsabs.harvard.edu/abs/2009arXiv0912.0201L} {} (\mn@eprint {arXiv}
  {0912.0201})

\bibitem[\protect\citeauthoryear{{Laureijs} et~al.,}{{Laureijs}
  et~al.}{2011}]{Euclid}
{Laureijs} R.,  et~al., 2011, preprint, \href
  {http://adsabs.harvard.edu/abs/2011arXiv1110.3193L} {} (\mn@eprint {arXiv}
  {1110.3193})

\bibitem[\protect\citeauthoryear{{Lawrence} et~al.,}{{Lawrence}
  et~al.}{2007}]{UKIDSS}
{Lawrence} A.,  et~al., 2007, \mn@doi [\mnras]
  {10.1111/j.1365-2966.2007.12040.x}, \href
  {https://ui.adsabs.harvard.edu/abs/2007MNRAS.379.1599L} {379, 1599}

\bibitem[\protect\citeauthoryear{{Leauthaud} et~al.,}{{Leauthaud}
  et~al.}{2012}]{Leauthaud12}
{Leauthaud} A.,  et~al., 2012, \mn@doi [\apj] {10.1088/0004-637X/744/2/159},
  \href {https://ui.adsabs.harvard.edu/abs/2012ApJ...744..159L} {744, 159}

\bibitem[\protect\citeauthoryear{{Linke} et~al.,}{{Linke}
  et~al.}{2020}]{Linke20}
{Linke} L.,  et~al., 2020, \mn@doi [\aap] {10.1051/0004-6361/202038355}, \href
  {https://ui.adsabs.harvard.edu/abs/2020A&A...640A..59L} {640, A59}

\bibitem[\protect\citeauthoryear{{Liske} et~al.,}{{Liske}
  et~al.}{2015}]{GAMA-II}
{Liske} J.,  et~al., 2015, \mn@doi [\mnras] {10.1093/mnras/stv1436}, \href
  {http://adsabs.harvard.edu/abs/2015MNRAS.452.2087L} {452, 2087}

\bibitem[\protect\citeauthoryear{{Mandelbaum}, {Wang}, {Zu}, {White},
  {Henriques}  \& {More}}{{Mandelbaum} et~al.}{2016}]{Mandelbaum16}
{Mandelbaum} R.,  {Wang} W.,  {Zu} Y.,  {White} S.,  {Henriques} B.,   {More}
  S.,  2016, \mn@doi [\mnras] {10.1093/mnras/stw188}, \href
  {https://ui.adsabs.harvard.edu/abs/2016MNRAS.457.3200M} {457, 3200}

\bibitem[\protect\citeauthoryear{{McFarland}, {Verdoes-Kleijn}, {Sikkema},
  {Helmich}, {Boxhoorn}  \& {Valentijn}}{{McFarland} et~al.}{2013}]{AW}
{McFarland} J.~P.,  {Verdoes-Kleijn} G.,  {Sikkema} G.,  {Helmich} E.~M.,
  {Boxhoorn} D.~R.,   {Valentijn} E.~A.,  2013, \mn@doi [Experimental
  Astronomy] {10.1007/s10686-011-9266-x}, \href
  {https://ui.adsabs.harvard.edu/abs/2013ExA....35...45M} {35, 45}

\bibitem[\protect\citeauthoryear{{Mead} \& {Verde}}{{Mead} \&
  {Verde}}{2021}]{MV20}
{Mead} A.~J.,  {Verde} L.,  2021, \mn@doi [\mnras] {10.1093/mnras/stab748},
  \href {https://ui.adsabs.harvard.edu/abs/2021MNRAS.503.3095M} {503, 3095}

\bibitem[\protect\citeauthoryear{{Miller} et~al.,}{{Miller}
  et~al.}{2013}]{lensfit}
{Miller} L.,  et~al., 2013, \mn@doi [\mnras] {10.1093/mnras/sts454}, \href
  {http://adsabs.harvard.edu/abs/2013MNRAS.429.2858M} {429, 2858}

\bibitem[\protect\citeauthoryear{{Nakoneczny} et~al.,}{{Nakoneczny}
  et~al.}{2021}]{Nakoneczny20}
{Nakoneczny} S.~J.,  et~al., 2021, \mn@doi [\aap]
  {10.1051/0004-6361/202039684}, \href
  {https://ui.adsabs.harvard.edu/abs/2021A&A...649A..81N} {649, A81}

\bibitem[\protect\citeauthoryear{{Navarro}, {Frenk}  \& {White}}{{Navarro}
  et~al.}{1997}]{NFW97}
{Navarro} J.~F.,  {Frenk} C.~S.,   {White} S. D.~M.,  1997, \mn@doi [\apj]
  {10.1086/304888}, \href
  {https://ui.adsabs.harvard.edu/abs/1997ApJ...490..493N} {490, 493}

\bibitem[\protect\citeauthoryear{{Pasquet}, {Bertin}, {Treyer}, {Arnouts}  \&
  {Fouchez}}{{Pasquet} et~al.}{2019}]{Pasquet19}
{Pasquet} J.,  {Bertin} E.,  {Treyer} M.,  {Arnouts} S.,   {Fouchez} D.,  2019,
  \mn@doi [\aap] {10.1051/0004-6361/201833617}, \href
  {https://ui.adsabs.harvard.edu/abs/2019A&A...621A..26P} {621, A26}

\bibitem[\protect\citeauthoryear{{Peacock} \& {Bilicki}}{{Peacock} \&
  {Bilicki}}{2018}]{PB18}
{Peacock} J.~A.,  {Bilicki} M.,  2018, \mn@doi [\mnras]
  {10.1093/mnras/sty2314}, \href
  {https://ui.adsabs.harvard.edu/abs/2018MNRAS.481.1133P} {481, 1133}

\bibitem[\protect\citeauthoryear{{Percival} et~al.,}{{Percival}
  et~al.}{2001}]{Percival01}
{Percival} W.~J.,  et~al., 2001, \mn@doi [\mnras]
  {10.1046/j.1365-8711.2001.04827.x}, \href
  {https://ui.adsabs.harvard.edu/abs/2001MNRAS.327.1297P} {327, 1297}

\bibitem[\protect\citeauthoryear{{Petrosian}}{{Petrosian}}{1976}]{Petro}
{Petrosian} V.,  1976, \mn@doi [\apjl] {10.1086/182301}, \href
  {https://ui.adsabs.harvard.edu/abs/1976ApJ...209L...1P} {210, L53}

\bibitem[\protect\citeauthoryear{{Porredon} et~al.,}{{Porredon}
  et~al.}{2021}]{Porredon20}
{Porredon} A.,  et~al., 2021, \mn@doi [\prd] {10.1103/PhysRevD.103.043503},
  \href {https://ui.adsabs.harvard.edu/abs/2021PhRvD.103d3503P} {103, 043503}

\bibitem[\protect\citeauthoryear{{Richard} et~al.,}{{Richard}
  et~al.}{2019}]{4MOST-CRS}
{Richard} J.,  et~al., 2019, \mn@doi [The Messenger] {10.18727/0722-6691/5127},
  \href {https://ui.adsabs.harvard.edu/abs/2019Msngr.175...50R} {175, 50}

\bibitem[\protect\citeauthoryear{{Robotham} et~al.,}{{Robotham}
  et~al.}{2011}]{Robotham11}
{Robotham} A.~S.~G.,  et~al., 2011, \mn@doi [\mnras]
  {10.1111/j.1365-2966.2011.19217.x}, \href
  {https://ui.adsabs.harvard.edu/abs/2011MNRAS.416.2640R} {416, 2640}

\bibitem[\protect\citeauthoryear{{Rozo} et~al.,}{{Rozo}
  et~al.}{2016}]{redMaGiC}
{Rozo} E.,  et~al., 2016, \mn@doi [\mnras] {10.1093/mnras/stw1281}, \href
  {http://adsabs.harvard.edu/abs/2016MNRAS.461.1431R} {461, 1431}

\bibitem[\protect\citeauthoryear{{Sadeh}, {Abdalla}  \& {Lahav}}{{Sadeh}
  et~al.}{2016}]{ANNz2}
{Sadeh} I.,  {Abdalla} F.~B.,   {Lahav} O.,  2016, \mn@doi [\pasp]
  {10.1088/1538-3873/128/968/104502}, \href
  {http://adsabs.harvard.edu/abs/2016PASP..128j4502S} {128, 104502}

\bibitem[\protect\citeauthoryear{{Schlafly} \& {Finkbeiner}}{{Schlafly} \&
  {Finkbeiner}}{2011}]{SF11}
{Schlafly} E.~F.,  {Finkbeiner} D.~P.,  2011, \mn@doi [\apj]
  {10.1088/0004-637X/737/2/103}, \href
  {https://ui.adsabs.harvard.edu/abs/2011ApJ...737..103S} {737, 103}

\bibitem[\protect\citeauthoryear{{Schlegel}, {Finkbeiner}  \&
  {Davis}}{{Schlegel} et~al.}{1998}]{SFD}
{Schlegel} D.~J.,  {Finkbeiner} D.~P.,   {Davis} M.,  1998, \mn@doi [\apj]
  {10.1086/305772}, \href {http://adsabs.harvard.edu/abs/1998ApJ...500..525S}
  {500, 525}

\bibitem[\protect\citeauthoryear{{Seljak}}{{Seljak}}{2000}]{Seljak00}
{Seljak} U.,  2000, \mn@doi [\mnras] {10.1046/j.1365-8711.2000.03715.x}, \href
  {https://ui.adsabs.harvard.edu/abs/2000MNRAS.318..203S} {318, 203}

\bibitem[\protect\citeauthoryear{{Sif{\'o}n} et~al.,}{{Sif{\'o}n}
  et~al.}{2015}]{Sifon15}
{Sif{\'o}n} C.,  et~al., 2015, \mn@doi [\mnras] {10.1093/mnras/stv2051}, \href
  {http://adsabs.harvard.edu/abs/2015MNRAS.454.3938S} {454, 3938}

\bibitem[\protect\citeauthoryear{{Singh}, {Mandelbaum}, {Seljak}, {Slosar}  \&
  {Vazquez Gonzalez}}{{Singh} et~al.}{2017}]{Singh17}
{Singh} S.,  {Mandelbaum} R.,  {Seljak} U.,  {Slosar} A.,   {Vazquez Gonzalez}
  J.,  2017, \mn@doi [\mnras] {10.1093/mnras/stx1828}, \href
  {https://ui.adsabs.harvard.edu/abs/2017MNRAS.471.3827S} {471, 3827}

\bibitem[\protect\citeauthoryear{{Soo} et~al.,}{{Soo} et~al.}{2018}]{Soo18}
{Soo} J.~Y.~H.,  et~al., 2018, \mn@doi [\mnras] {10.1093/mnras/stx3201}, \href
  {http://adsabs.harvard.edu/abs/2018MNRAS.475.3613S} {475, 3613}

\bibitem[\protect\citeauthoryear{{Strauss} et~al.,}{{Strauss}
  et~al.}{2002}]{SDSS.MGS}
{Strauss} M.~A.,  et~al., 2002, \mn@doi [\aj] {10.1086/342343}, \href
  {https://ui.adsabs.harvard.edu/abs/2002AJ....124.1810S} {124, 1810}

\bibitem[\protect\citeauthoryear{{Sugiyama}, {Takada}, {Kobayashi}, {Miyatake},
  {Shirasaki}, {Nishimichi}  \& {Park}}{{Sugiyama} et~al.}{2020}]{Sugiyama20}
{Sugiyama} S.,  {Takada} M.,  {Kobayashi} Y.,  {Miyatake} H.,  {Shirasaki} M.,
  {Nishimichi} T.,   {Park} Y.,  2020, \mn@doi [\prd]
  {10.1103/PhysRevD.102.083520}, \href
  {https://ui.adsabs.harvard.edu/abs/2020PhRvD.102h3520S} {102, 083520}

\bibitem[\protect\citeauthoryear{{Taylor}}{{Taylor}}{2005}]{TOPCAT}
{Taylor} M.~B.,  2005, in {Shopbell} P.,  {Britton} M.,   {Ebert} R.,  eds,
  Astronomical Society of the Pacific Conference Series Vol. 347, Astronomical
  Data Analysis Software and Systems XIV. p.~29

\bibitem[\protect\citeauthoryear{{Taylor}}{{Taylor}}{2006}]{STILTS}
{Taylor} M.~B.,  2006, in {Gabriel} C.,  {Arviset} C.,  {Ponz} D.,   {Enrique}
  S.,  eds,  Astronomical Society of the Pacific Conference Series Vol. 351,
  Astronomical Data Analysis Software and Systems XV. p.~666

\bibitem[\protect\citeauthoryear{{Taylor} et~al.,}{{Taylor}
  et~al.}{2011}]{Taylor11}
{Taylor} E.~N.,  et~al., 2011, \mn@doi [\mnras]
  {10.1111/j.1365-2966.2011.19536.x}, \href
  {https://ui.adsabs.harvard.edu/abs/2011MNRAS.418.1587T} {418, 1587}

\bibitem[\protect\citeauthoryear{{The Dark Energy Survey Collaboration}}{{The
  Dark Energy Survey Collaboration}}{2005}]{DES}
{The Dark Energy Survey Collaboration} 2005, preprint, \href
  {http://adsabs.harvard.edu/abs/2005astro.ph.10346T} {} (\mn@eprint {arXiv}
  {astro-ph/0510346})

\bibitem[\protect\citeauthoryear{{Tian}, {Hoekstra}  \& {Zhao}}{{Tian}
  et~al.}{2009}]{Tian09}
{Tian} L.,  {Hoekstra} H.,   {Zhao} H.,  2009, \mn@doi [\mnras]
  {10.1111/j.1365-2966.2008.14094.x}, \href
  {https://ui.adsabs.harvard.edu/abs/2009MNRAS.393..885T} {393, 885}

\bibitem[\protect\citeauthoryear{Tinker, Robertson, Kravtsov, Klypin, Warren,
  Yepes  \& Gottl{\"{o}}ber}{Tinker et~al.}{2010}]{Tinker10}
Tinker J.~L.,  Robertson B.~E.,  Kravtsov A.~V.,  Klypin A.,  Warren M.~S.,
  Yepes G.,   Gottl{\"{o}}ber S.,  2010, \mn@doi [Astrophys. J.]
  {10.1088/0004-637X/724/2/878}, 724, 878

\bibitem[\protect\citeauthoryear{{Tr{\"o}ster} et~al.,}{{Tr{\"o}ster}
  et~al.}{2021}]{Troster20}
{Tr{\"o}ster} T.,  et~al., 2021, \mn@doi [\aap] {10.1051/0004-6361/202039805},
  \href {https://ui.adsabs.harvard.edu/abs/2021A&A...649A..88T} {649, A88}

\bibitem[\protect\citeauthoryear{{Vakili} et~al.,}{{Vakili}
  et~al.}{2019}]{Vakili19}
{Vakili} M.,  et~al., 2019, \mn@doi [\mnras] {10.1093/mnras/stz1249}, \href
  {https://ui.adsabs.harvard.edu/abs/2019MNRAS.487.3715V} {487, 3715}

\bibitem[\protect\citeauthoryear{{Vakili} et~al.,}{{Vakili}
  et~al.}{2020}]{Vakili20}
{Vakili} M.,  et~al., 2020, preprint, \href
  {https://ui.adsabs.harvard.edu/abs/2020arXiv200813154V} {} (\mn@eprint
  {arXiv} {2008.13154})

\bibitem[\protect\citeauthoryear{{Velander} et~al.,}{{Velander}
  et~al.}{2014}]{Velander14}
{Velander} M.,  et~al., 2014, \mn@doi [\mnras] {10.1093/mnras/stt2013}, \href
  {http://adsabs.harvard.edu/abs/2014MNRAS.437.2111V} {437, 2111}

\bibitem[\protect\citeauthoryear{{Viola} et~al.,}{{Viola}
  et~al.}{2015}]{Viola15}
{Viola} M.,  et~al., 2015, \mn@doi [\mnras] {10.1093/mnras/stv1447}, \href
  {http://adsabs.harvard.edu/abs/2015MNRAS.452.3529V} {452, 3529}

\bibitem[\protect\citeauthoryear{{Wechsler} \& {Tinker}}{{Wechsler} \&
  {Tinker}}{2018}]{Wechsler18}
{Wechsler} R.~H.,  {Tinker} J.~L.,  2018, \mn@doi [\araa]
  {10.1146/annurev-astro-081817-051756}, \href
  {https://ui.adsabs.harvard.edu/abs/2018ARA&A..56..435W} {56, 435}

\bibitem[\protect\citeauthoryear{{Wolf} et~al.,}{{Wolf}
  et~al.}{2017}]{2dFLenS-photo-z}
{Wolf} C.,  et~al., 2017, \mn@doi [\mnras] {10.1093/mnras/stw3151}, \href
  {http://adsabs.harvard.edu/abs/2017MNRAS.466.1582W} {466, 1582}

\bibitem[\protect\citeauthoryear{{Wright}}{{Wright}}{2016}]{LAMBDAR-software}
{Wright} A.~H.,  2016, {LAMBDAR: Lambda Adaptive Multi-Band Deblending
  Algorithm in R}, Astrophysics Source Code Library (\mn@eprint {ascl}
  {1604.003})

\bibitem[\protect\citeauthoryear{{Wright} et~al.,}{{Wright}
  et~al.}{2016}]{GAMA-LAMBDAR}
{Wright} A.~H.,  et~al., 2016, \mn@doi [\mnras] {10.1093/mnras/stw832}, \href
  {http://adsabs.harvard.edu/abs/2016MNRAS.460..765W} {460, 765}

\bibitem[\protect\citeauthoryear{{Wright} et~al.,}{{Wright}
  et~al.}{2019}]{KV450-data}
{Wright} A.~H.,  et~al., 2019, \mn@doi [\aap] {10.1051/0004-6361/201834879},
  \href {https://ui.adsabs.harvard.edu/abs/2019A&A...632A..34W} {632, A34}

\bibitem[\protect\citeauthoryear{{Wright}, {Hildebrandt}, {van den Busch},
  {Heymans}, {Joachimi}, {Kannawadi}  \& {Kuijken}}{{Wright}
  et~al.}{2020}]{KV450-SOM}
{Wright} A.~H.,  {Hildebrandt} H.,  {van den Busch} J.~L.,  {Heymans} C.,
  {Joachimi} B.,  {Kannawadi} A.,   {Kuijken} K.,  2020, \mn@doi [\aap]
  {10.1051/0004-6361/202038389}, \href
  {https://ui.adsabs.harvard.edu/abs/2020A&A...640L..14W} {640, L14}

\bibitem[\protect\citeauthoryear{Yang, Mo  \& van~den Bosch}{Yang
  et~al.}{2008}]{Yang08}
Yang X.,  Mo H.~J.,   van~den Bosch F.~C.,  2008, \mn@doi [Astrophys. J.]
  {10.1086/528954}, 676, 248

\bibitem[\protect\citeauthoryear{{de Jong}, {Verdoes Kleijn}, {Kuijken}  \&
  {Valentijn}}{{de Jong} et~al.}{2013}]{KiDS}
{de Jong} J.~T.~A.,  {Verdoes Kleijn} G.~A.,  {Kuijken} K.~H.,   {Valentijn}
  E.~A.,  2013, \mn@doi [Experimental Astronomy] {10.1007/s10686-012-9306-1},
  \href {http://adsabs.harvard.edu/abs/2013ExA....35...25D} {35, 25}

\bibitem[\protect\citeauthoryear{{de Jong} et~al.,}{{de Jong}
  et~al.}{2015}]{KiDS-DR2}
{de Jong} J.~T.~A.,  et~al., 2015, \mn@doi [\aap]
  {10.1051/0004-6361/201526601}, \href
  {http://adsabs.harvard.edu/abs/2015A%26A...582A..62D} {582, A62}

\bibitem[\protect\citeauthoryear{{de Jong} et~al.,}{{de Jong}
  et~al.}{2017}]{KiDS-DR3}
{de Jong} J.~T.~A.,  et~al., 2017, \mn@doi [\aap]
  {10.1051/0004-6361/201730747}, \href
  {http://adsabs.harvard.edu/abs/2017A%26A...604A.134D} {604, A134}

\bibitem[\protect\citeauthoryear{{de Jong} et~al.,}{{de Jong}
  et~al.}{2019}]{4MOST}
{de Jong} R.~S.,  et~al., 2019, \mn@doi [The Messenger]
  {10.18727/0722-6691/5117}, \href
  {https://ui.adsabs.harvard.edu/abs/2019Msngr.175....3D} {175, 3}

\bibitem[\protect\citeauthoryear{{van Uitert} et~al.,}{{van Uitert}
  et~al.}{2016}]{vanUitert16}
{van Uitert} E.,  et~al., 2016, \mn@doi [\mnras] {10.1093/mnras/stw747}, \href
  {http://adsabs.harvard.edu/abs/2016MNRAS.459.3251V} {459, 3251}

\bibitem[\protect\citeauthoryear{{van Uitert} et~al.,}{{van Uitert}
  et~al.}{2017}]{vanUitert17}
{van Uitert} E.,  et~al., 2017, \mn@doi [\mnras] {10.1093/mnras/stx344}, \href
  {http://adsabs.harvard.edu/abs/2017MNRAS.467.4131V} {467, 4131}

\bibitem[\protect\citeauthoryear{{van Uitert} et~al.,}{{van Uitert}
  et~al.}{2018}]{vanUitert18}
{van Uitert} E.,  et~al., 2018, \mn@doi [\mnras] {10.1093/mnras/sty551}, \href
  {http://adsabs.harvard.edu/abs/2018MNRAS.476.4662V} {476, 4662}

\bibitem[\protect\citeauthoryear{{van den Busch} et~al.,}{{van den Busch}
  et~al.}{2020}]{KiDS-mocks}
{van den Busch} J.~L.,  et~al., 2020, \mn@doi [\aap]
  {10.1051/0004-6361/202038835}, \href
  {https://ui.adsabs.harvard.edu/abs/2020A&A...642A.200V} {642, A200}

\bibitem[\protect\citeauthoryear{van~der Walt, Colbert  \& Varoquaux}{van~der
  Walt et~al.}{2011}]{NumPy}
van~der Walt S.,  Colbert S.~C.,   Varoquaux G.,  2011, \mn@doi [Computing in
  Science Engineering] {10.1109/MCSE.2011.37}, 13, 22

\makeatother
\end{thebibliography}


\onecolumn
\begin{appendix}

\section{Dependence of photometric redshift quality on survey systematics}
\label{App: Systematics}

\begin{figure*}
    \centering
    \includegraphics[width=0.75\textwidth]{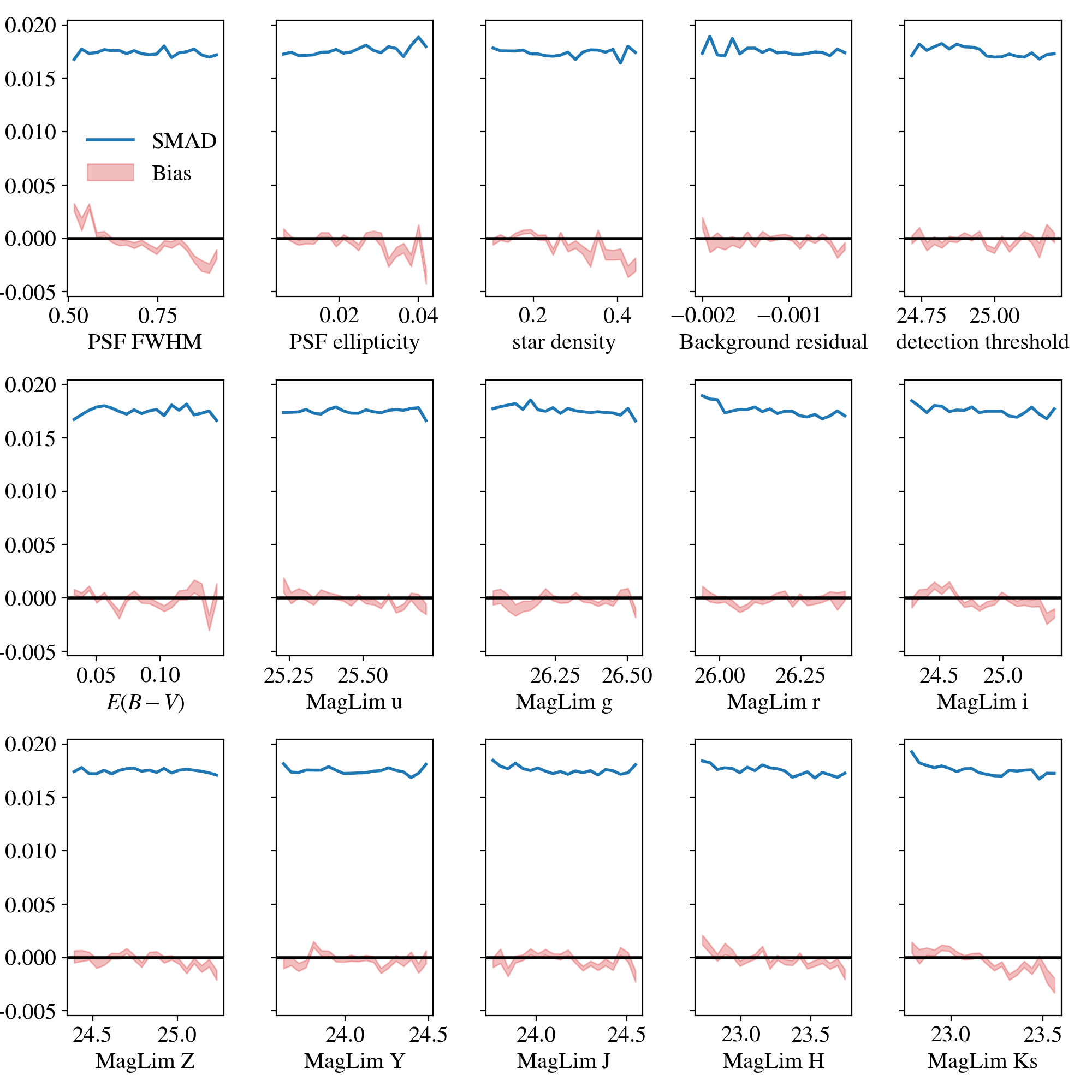}
    \caption{Photometric redshift quality (bias and scatter) as a function of KiDS-internal and external observational properties.}
    \label{Fig:photo-z-systematics}
\end{figure*}

Here we present how the \phzs\ of the KiDS-Bright sample described in Sect.~\ref{Sec:photo-zs} vary as a function of survey-related effects. In Figure \ref{Fig:photo-z-systematics} we show the \phz\ bias and scatter (SMAD) evaluated for a range of the following parameters:
\begin{itemize}
    \item \textbf{PSF FWHM} (full width at half maximum) in the $r$-band, in units  of  arcseconds, calculated using the \ttt{PSF\_Strehl\_ratio} column in the catalog;
    \item \textbf{PSF ellipticity} in the $r$-band, obtained from the \ttt{PSFe1} and \ttt{PSFe2} columns;
    \item \textbf{Star density} (projected), determined from  the  pixelated  number  density  map  of  bright stars  in  the  second  Gaia data  release \citep{Gaia-DR2};
    \item \textbf{Background residual} counts in the centroid positions of  the  objects  in  the  THELI-processed $r$-band  detection  images, provided  as \ttt{BACKGROUND} in the catalog; 
    \item \textbf{Detection  threshold}  above  background in  units  of  counts, provided as \ttt{THRESHOLD};
    \item \textbf{E(B-V)}, Galactic dust extinction in the $r$-band, derived from the \cite{SFD} maps with the \cite{SF11} corrections, provided as \ttt{EXTINCTION\_r} in the catalog;
and    \item \textbf{MagLim}, limiting magnitudes in the 9 KV bands, evaluated at object position.
\end{itemize}
For more details on these quantities, please see \cite{Vakili20}.

\section{Halo model}
\label{App:hod}

We modeled the halo occupation statistics using the conditional stellar mass function \citep[CSMF, as also presented by][]{Yang08,Cacciato13,vanUitert16}, and we employed them to calculate the $\mathcal{H}$ functions used in the halo model \citep{Cacciato13, vanUitert16, Dvornik18}. The CSMF, $\Phi(M_{\star} \vert M_{\mathrm{h}})$, specifies the average number of galaxies of stellar mass $M_{\star}$ that reside in a halo of mass $M_{\mathrm{h}}$. In this formalism, the halo occupation statistics of central galaxies are defined via the following function: 
\begin{equation}\label{CLFsplit}
\Phi(M_{\star} \vert M_{\mathrm{h}}) = \Phi_{\mathrm{c}}(M_{\star}  \vert M_{\mathrm{h}}) + \Phi_{\mathrm{s}}(M_{\star}  \vert M_{\mathrm{h}})\,.
\end{equation}
In particular, the CSMF of central galaxies was modeled as a log-normal,
\begin{equation}\label{phi_c}
\Phi_{\mathrm{c}}(M_{\star}  \vert M_{\mathrm{h}}) = {1 \over {\sqrt{2\pi} \, {\ln}(10)\, \sigma_{\text{c}} M_{\star} } 
}{\exp}\left[- { {\log(M_{\star} / M^{*}_{\mathrm{c}} )^2 } \over 2\,\sigma_{\mathrm{c}}^{2}} \right]\, \,,
\end{equation}
and the satellite term as a modified Schechter function,
\begin{equation}\label{phi_s}
\Phi_{\mathrm{s}}(M_{\star}  \vert M_{\mathrm{h}}) = { \phi^{*}_{\mathrm{s}} \over M^{*}_{\mathrm{s}}}\,
\left({M_{\star} \over M^{*}_{\mathrm{s}}}\right)^{\alpha_{\mathrm{s}}} \,
{\exp} \left[- \left ({M_{\star} \over M^{*}_{\text{s}}}\right )^2 \right] 
\,,
\end{equation}
where $\sigma_{\text{c}}$ is the scatter between stellar mass and halo mass and $\alpha_{\text{s}}$ governs the power law behavior of satellite galaxies.  We note that $M^{*}_{\text{c}}$, $\sigma_{\text{c}}$, $\phi^{*}_{\text{s}}$, $\alpha_{\text{s}}$, and
$M^{*}_{\text{s}}$ are, in principle, all functions of halo mass $M_\mrm{h}$. We assume that $\sigma_{\text{c}}$ and $\alpha_{\text{s}}$ are independent of the halo mass $M_{\mathrm{h}}$. Halo masses are drawn from the halo mass function for which we assume the \citet{Tinker10} fitting function. Inspired by \citet{Yang08}, we parameterize $M^{*}_{\text{c}}$, $M^{*}_{\text{s}}$, and $\phi^{*}_{\text{s}}$ as:
\begin{equation}\label{eq:3_CMF4}
M^{*}_{\mathrm{c}}(M_{\mathrm{h}}) = M_{0} \frac{(M_{\mathrm{h}}/M_{1})^{\gamma_{1}}}{[1 + (M_{\mathrm{h}}/M_{1})]^{\gamma_{1} - \gamma_{2}}}\,,\end{equation}
\begin{equation}\label{eq:3_CMF5}
M_{\mathrm{s}}^{*}(M_{\mathrm{h}}) = 0.56\ M^{*}_{\mathrm{c}}(M_{\mathrm{h}})\,,
\end{equation}
and
\begin{equation}\label{eq:3_CMF7}
\log[\phi_{\mathrm{s}}^{*}(M_{\mathrm{h}})] = b_{0} + b_{1}(\log m_{13})\,,
\end{equation}
where $m_{13} = M_{\mathrm{h}}/(10^{13}M_{\odot})$. The factor of $0.56$ is also inspired by \citet{Yang08} and further tests by \citet{vanUitert16} have shown that using this assumption does not significantly affect the results.

From the CSMF, it is straightforward to compute the halo occupation numbers.  The average number of galaxies with stellar masses in the range from $M_{\star,1} \leq M_{\star}  \leq M_{\star,2}$ is thus given by the following:
\begin{equation}\label{HODfromCLF}
\langle N_{x} \vert M_{\mathrm{h}} \rangle = \int_{M_{\star,1}}^{M_{\star,2}} \Phi_{x}(M_{\star}  \vert M_{\mathrm{h}}) \, \mathrm{d} M_{\star} \,,
\end{equation}
where $x$ stands for either central or satellite. For the two components, we can then write
\begin{equation}
\mathcal{H}_{x}(k, M_{\mathrm{h}}) = {\langle N_{x} \vert M_{\mathrm{h}} \rangle \over \overline{n}_{x}} \,  \tilde{u}_{x}(k \vert M_{\mathrm{h}})\,,
\end{equation}
where $\tilde{u}_{x}(k \vert M_{\mathrm{h}})$ are the normalized Fourier transforms of the radial distribution of the central or satellite galaxies. For centrals, we assume that $\tilde{u}_{x}(k \vert M_{\mathrm{h}}) = 1$ and for satellites $\tilde{u}_{x}(k \vert M_{\mathrm{h}}) = \tilde{u}_{\mathrm{h}}(k \vert M_{\mathrm{h}})$ (the satellite distribution follows the dark matter). The average number density $\overline{n}_{x}$ follows from
\begin{equation}\label{averng}
\overline{n}_{x} = \int_{0}^{\infty} \langle N_{x} \vert M_{\mathrm{h}} \rangle \, n(M_{\mathrm{h}}) \, \mathrm{d} M_{\mathrm{h}}\,,
\end{equation}
where $n(M_{\mathrm{h}})$ is the halo mass function. For the dark matter we have the following:
\begin{equation}
\mathcal{H}_{\mathrm{m}}(k, M_{\mathrm{h}}) = {M_{\mathrm{h}} \over \overline{\rho}_{\mathrm{m}}} \,  \tilde{u}_{\mathrm{h}}(k \vert M_{\mathrm{h}})\,,
\end{equation}
where $\overline{\rho}_{\text{m}}$ is the mean density of the Universe and $\tilde{u}_{\text{h}}(k \vert M_{\mathrm{h}})$ is the normalized Fourier transform of the NFW profile \citep{NFW97}. Using these ingredients, one can construct the following 1-halo and 2-halo power spectra \citep[see also Equations 5 -- 7 in][]{vanUitert16}:
\begin{equation}
P^{\mathrm{1h}}_{xy}(k) = \int_{0}^{\infty} \mathcal{H}_{x}(k, M_{\mathrm{h}}) \, \mathcal{H}_{y}(k, M_{\mathrm{h}}) \, n(M_{\mathrm{h}}) \, \mathrm{d} M_{\mathrm{h}}\,\end{equation}
and
\begin{equation}
{P^{\mathrm{2h}}_{xy}(k) = P_{\text{lin}}(k) \,  \int_{0}^{\infty} \mathrm{d} M_{\mathrm{h},1} \, \mathcal{H}_{x}(k, M_{\mathrm{h},1}) \, b_{\mathrm{h}}(M_{\mathrm{h},1})\, n(M_{\mathrm{h},1})}  
\, \int_{0}^{\infty} \mathrm{d} M_{\mathrm{h},2} \, \mathcal{H}_{y}(k, M_{\mathrm{h},2}) \, b_{\mathrm{h}}(M_{\mathrm{h},2})\, n(M_{\mathrm{h},2}) \,,
\end{equation}
where $b_{\mathrm{h}}(M_{\mathrm{h}})$ is the halo bias from \citet{Tinker10} and $P_{\text{lin}}(k)$ is the linear matter power spectrum. The full GGL power spectrum is thus written as $P_{\mathrm{gm}}(k) = P^{\mathrm{1h}}_{\mathrm{cm}}(k) + P^{\mathrm{1h}}_{\mathrm{sm}}(k) + P^{\mathrm{2h}}_{\mathrm{cm}}(k) + P^{\mathrm{2h}}_{\mathrm{sm}}(k)$, from which the $\Delta \Sigma_{\mathrm{gm}}$ can be calculated using the following Fourier and Abel transforms \citep[see also Equations 1 -- 4 of][]{vanUitert16}:
\begin{equation}
\xi_{\mathrm{gm}}(r) = {1 \over 2 \pi^{2}} \int_{0}^{\infty} P_{\mathrm{gm}}(k) \, {\sin kr \over kr} \, k^{2} \, \mathrm{d} k\,, 
\end{equation}
\begin{equation}
\Sigma_{\mathrm{gm}}(r_{\mathrm{p}}) = 2  {\overline{\rho}_{\mathrm{m}}} \, \int_{r_{\mathrm{p}}}^{\infty} \xi_{\mathrm{gm}}(r) \, { r\, \mathrm{d} r \over \sqrt{r^2 - r_{\mathrm{p}}^2}}\,,
\end{equation}
where $r_{\mathrm{p}}$ is the projected separation.  We also define $\overline{\Sigma}_{\mathrm{xy}}(< r_{\mathrm{p}})$ as its average inside $r_{\mathrm{p}}$:
\begin{equation}
\overline{\Sigma}_{\mathrm{gm}}(< r_{\mathrm{p}})  = \frac{2}{r^2_{\mathrm{p}}} \int_{0}^{r_{\mathrm{p}}}\Sigma_{\mathrm{gm}}(R') R'\, \mathrm{d}R'\,,
\end{equation}
which we used to define the excess surface density (ESD)
\begin{equation}
\Delta \Sigma_{\mathrm{gm}}(r_{\mathrm{p}}) = \overline{\Sigma}_{\mathrm{gm}}(< r_{\mathrm{p}}) - \Sigma_{\mathrm{gm}}(r_{\mathrm{p}})\,.
\end{equation}
For completeness, we included the contribution of the stellar mass of galaxies to the lensing signal as a point mass, so that $\Delta \Sigma^{\mathrm{pm}}_{\mathrm{gm}}(r_{\mathrm{p}}) = M_{\star, \mathrm{med}} / \pi r_{\mathrm{p}}^{2}$.

\section{Details of released data}
\label{App:data_release}

Here we provide a description of the columns for the KiDS-1000 bright galaxy sample data release available at \url{http://kids.strw.leidenuniv.nl/DR4/brightsample.php}. It is separated into the photometric redshift catalog and the \textsc{LePHARE} derivations. The catalogs can be cross-matched by ID between each other and with the KiDS Data Release 4 main dataset available from \url{http://kids.strw.leidenuniv.nl/DR4/index.php}.

We provide a list of columns contained in the photometric redshift catalog below:

\begin{itemize}

\item  \ttt{ID}: source identifier from the KiDS DR4 catalog.

\item \ttt{RAJ2000}: right ascension (J2000).

\item \ttt{DECJ2000}: declination (J2000).

\item \ttt{MAG\_AUTO\_calib}: zero-point calibrated and extinction-corrected Kron-like elliptical aperture magnitude in the $r$ band; $\mtt{MAG\_AUTO\_calib} = \mtt{MAG\_AUTO} + \mtt{DMAG} - \mtt{EXTINCTION\_R}$.

\item \ttt{MAGERR\_AUTO}: RMS error for MAG\_AUTO.

\item \ttt{zphot\_ANNz2}: photometric redshift derived with ANNz2.

\item \ttt{MASK}: 9-band mask information.

\item \ttt{masked}: binary flag, set to 0 for unmasked and to 1 for masked objects. Use $\mtt{masked}==0$ for the default selection.

\end{itemize}

We also provide a list of columns contained in the stellar mass catalog below:

\begin{itemize}

        \item \ttt{ID}: source identifier from the KiDS DR4 catalog.
        
        \item \ttt{RAJ2000}: right ascension (J2000).

\item \ttt{DECJ2000}: declination (J2000).

    \item     \ttt{K\_COR\_x}: the K-correction for the x-band.

    \item     \ttt{MAG\_ABS\_x}: the absolute magnitude in the x-band.

    \item     \ttt{MABS\_FILTx}: the filter that is used for reference when computing the MABS.

    \item     \ttt{CONTEXT}: a bit flag that shows which photometric bands were used in the fitting process. That is to say, if 9-band information was employed, the bit flag is as follows: 111111111=1+2+4+8+16+32+64+128+256=511. If a $Z$-band is missing, then the bit flag is as follows: 111101111=1+2+4+8+0+32+64+128+256=495.

    \item     \ttt{REDSHIFT}: the redshift values used for the stellar mass computation, in this case \phzs\ derived with ANNz2. 
    
    \item    \ttt{MASS\_MED}: the median of the galaxy template stellar mass PDF measured by \textsc{LePHARE}. It is important to note the galaxies with MASS\_MED == -99 were best-fit by a non-galaxy template, but the  MASS\_BEST value still shows the best fitting galaxy template mass for them, nonetheless.
   
    \item     \ttt{MASS\_INF}: the lower-limit on the stellar mass from the galaxy mass PDF (68\% confidence level).
    
    \item     \ttt{MASS\_SUP}: the upper-limit on the stellar mass from the galaxy mass PDF (68\% confidence level).

    \item     \ttt{MASS\_BEST}: the best-fit stellar mass estimated by \textsc{LePHARE}. This column should be used as the stellar mass, but it is necessary to make sure to apply the fluxscale correction (see below).

    \item     \ttt{SFR\_INF}: the lower-limit on the star formation rate from the galaxy SFR PDF (68\% confidence level).

    \item 
    \ttt{SFR\_SUP}: the lower-limit on the star formation rate from the galaxy SFR PDF (68\% confidence level).

    \item 
    \ttt{SFR\_BEST}: best-fit star formation rate (SFR) estimated by \textsc{LePHARE}.

\end{itemize}
\textbf{Note 1.} All the "\ttt{MASS}" quantities stand for $\log\!10(M_\star / M_\sun)$.\\
\textbf{Note 2.} Fluxscale correction: Because the \textsc{GAaP} photometry only measures the galaxy magnitude within a specific aperture size, the stellar mass should be corrected using a ``fluxscale'' parameter, which is the ratio of \ttt{AUTO} and \textsc{GAaP} fluxes:
\begin{equation}
\label{Eq:fluxscale}
    \log\!10(\mtt{fluxscale}) = (\mtt{MAG\_GAAP\_r} - \mtt{MAG\_AUTO}) / 2.5.
\end{equation}
The "total" stellar mass in then
\begin{equation}
\label{Eq:Mstar-tot}
    \mtt{M\_TOT} = \mtt{M\_BEST} + \log\!10(\mtt{fluxscale}).
\end{equation}
Similarly, also absolute magnitudes need corrections if total measurements are required:
\begin{equation}
\label{Eq:Abs-mag}
   \mtt{MAG\_ABS\_X,total} =  \mtt{MAG\_ABS\_X} - 2.5 \log\!10(\mtt{fluxscale}).
\end{equation}

All the \textsc{LePhare} quantities were computed assuming $h=0.7$, and the estimated stellar masses are assumed to have a dependence on $h$ dominated by the $h^{-2}$ scaling of luminosities. Therefore, if another Hubble constant value is used, the logarithmic stellar mass in Eq.~\eqref{Eq:Mstar-tot} needs to be corrected by $-2\log\!10 (h/0.7)$, while the absolute magnitudes in Eq.~\eqref{Eq:Abs-mag} need to have $5 \log\! 10 (h/0.7)$ added.

\end{appendix}


\end{document}